\documentclass[aps,showpacs,twocolumn,prd,preprintnumbers,amsmath,amssymb,letterpaper]{revtex4}
\newcommand{\pd}[2]{ { \partial {#1} \over \partial {#2} } }

\def\beqar{\begin{eqnarray}}
\def\eeqar{\end{eqnarray}}
\newcommand{\llabel}[1]{\label{#1}}              

\newcommand{\labeq}[2]{ \begin{equation} \llabel{#1}
{#2}
\end{equation}}

\newcommand{\beq}{\begin{equation}}
\newcommand{\eeq}{\end{equation}}
\usepackage{graphicx}
\usepackage{epsf}
\usepackage{amsmath}
\usepackage{graphics,epsfig,placeins,subfigure,wrapfig}
\begin{document}
\title{The merger of white dwarf-neutron star binaries: \\ Prelude to hydrodynamic simulations in general relativity}
\author{Vasileios Paschalidis${}^1$, Morgan MacLeod${}^2$, Thomas W. Baumgarte${}^{1,2}$, and Stuart L. Shapiro${}^{1,3}$}
%
\affiliation{${}^1$ Department of Physics, University of Illinois at Urbana-Champaign,
Urbana, IL 61801 \\  
${}^2$Department of Physics and Astronomy, Bowdoin College,
Brunswick, ME 04011 \\
${}^3$Department of Astronomy and NCSA, University of Illinois at Urbana-Champaign,
Urbana, IL 61801
}
\date{\today}
\begin{abstract}
White dwarf--neutron star binaries generate detectable gravitational radiation. We construct Newtonian equilibrium models of corotational white dwarf--neutron star (WDNS) binaries in circular orbit and find that these models terminate at the Roche limit.  At this point the binary will undergo either stable mass transfer (SMT) and evolve on a secular timescale, or unstable mass transfer (UMT), which results in the tidal disruption of the WD. The path a given binary will follow depends primarily on its mass ratio. We  analyze  the  fate of known WDNS binaries and use population synthesis results to estimate the number of LISA-resolved galactic binaries that will undergo either SMT or UMT. We model the quasistationary SMT epoch by solving a set of simple ordinary differential equations and compute the corresponding gravitational waveforms. Finally, we discuss in general terms the possible fate of binaries that undergo UMT and construct approximate Newtonian equilibrium configurations of merged WDNS remnants. We use these configurations to assess plausible outcomes of our future, fully relativistic simulations of these systems. If sufficient WD debris lands on the NS, the remnant may collapse, whereby the gravitational waves from the inspiral, merger, and collapse phases will sweep from LISA through LIGO frequency bands. If the debris forms a disk about the NS, it may fragment and form planets.
\end{abstract}


\maketitle
%
%
 
\section{Introduction}

The inspiral and merger of compact binaries represent some of the most promising sources of gravitational waves (GWs) for detection by ground-based laser interferometers like LIGO \cite{LIGO1,LIGO2}, VIRGO \cite{VIRGO1,VIRGO2}, GEO \cite{GEO}, TAMA \cite{TAMA1,TAMA2} and AIGO \cite{AIGO}, as well as by proposed space-based interferometers like LISA \cite{LISA} and DECIGO \cite{DECIGO}. Extracting physical information from gravitational radiation emitted by compact binaries requires careful modeling of these systems (see \cite{BSReview} for a review). Most effort to date has focused on modeling black hole--black hole (BHBH) binaries \cite{Pretorius2005a,Baker2006,Campanelli2006,Brugmann2006a,Herrmann2006,Sperhake2006,Etienne2007a,Healy,Baker2006b,Lousto2007,Gonzalez08,Scheel2008} and neutron star--neutron star (NSNS) binaries \cite{2004PhRvD..69l4036F,2005A&A...431..297B,2005PhRvD..71h4021S,2006PhRvD..73f4027S,2008PhRvD..77b4006A,2008PhRvD..78b4012L,2008PhRvD..78h4033B}, with some recent work 
on black hole--neutron star (BHNS) binaries \cite{Rantsiou08, Loffler06, Faber06, Shibata07, Shibata08,Yamamoto08, Etienne08a, Etienne08, Duez08}. 

In this paper we begin to explore WDNS binaries. They are plausible
sources of low-frequency GWs for LISA, DECIGO and possibly also, as we shall see, high-frequency GWs for LIGO, VIRGO, GEO, TAMA and AIGO. 

Like NSNS binaries, WDNS binaries are known to exist. We have compiled a list of 
observed WDNS binaries 
in Tables \ref{table1} and \ref{table2}. Table~\ref{table1} lists those binaries with relatively well-determined individual component masses. By ``relatively well-determined" we mean that in determining the masses no assumption was made about the mass of the NS. 
Table~\ref{table2} lists those binaries for which the masses of the individual components have not been well-determined as yet. For these cases the NSs have been assumed to have a mass of either $1.35 M_\odot$ or $1.40 M_\odot$.
Of all the objects presented in Tables \ref{table1} and \ref{table2} 
only B2303+46, J1141-6545, J0751+1807 and J1757-5322 have been positively identified as a pulsar with a WD companion. For all other systems there is at best strong evidence that the companion of the observed pulsar is a WD. 

The frequency of GWs emitted by the binary with the shortest period, J1141$-$6545, is $\simeq 1.2\times 10^{-4} \rm Hz$. This frequency lies in the expected band of LISA, but well below the cutoff  frequency of $\simeq 10^{-3} \rm Hz$ due to the double WD confusion background \cite{Nelemans01}.  It would therefore be impossible to resolve this signal.    Assuming the quadrupole approximation for a WDNS binary with masses $M_{\rm WD}$, $M_{\rm NS}$, reduced mass $\mu$, and total mass $M_{\rm T}=M_{\rm NS}+M_{\rm WD}$ in circular orbit with angular frequency $\Omega$, the amplitude of GWs coming from this object is $h = 4 \mu M_{\rm T} / (r A)$ \cite{Shapiro}, where $r$ is the distance to the object and $A$ the binary separation. In \cite{distanceJ11} a lower bound is given to the distance to PSR J1141$-$6545 of $r\geq 3.7$ kpc. Using the data of Table \ref{table1} we find that the maximum amplitude of the expected waves coming from this object is $h_{max}=1.36\times 10^{-23}$. A gravitational wave of this amplitude is below LISA's sensitivity at $10^{-4} \rm Hz$ ($h_s\sim 10^{-21}$
) and hence undetectable. In order for detectable gravitational wave signals to be emitted the orbital separation of PSR J1141$-$6545 has to decrease by a factor of about 100.

The emission of gravitational radiation will cause a binary to inspiral to a close, nearly circular orbit.  In this regime the binary can be described by a sequence of quasiequilibrium configurations.  Once the separation of a WDNS binary becomes a few times larger than the size of the WD, the amplitude of the emitted GWs, as we will see,  is large enough to allow for detection by space-based gravitational wave observatories. An important issue to address then is the number of Galactic WDNS binaries that could be detected by space-based interferometers.

\begin{table*}[t]
\caption{WDNS binaries with well-determined masses. From left to right the columns give the name of the object, the orbital period, the corresponding quadrupole GW frequency, the WD mass, the NS mass, the total mass, the mass ratio $q=M_{\rm WD}/M_{\rm NS}$ and the stability of mass transfer after its onset.}
\begin{tabular}{lccccccc}\hline\hline
\multicolumn{1}{p{2.15 cm}}{\hspace{0.3 cm} PSR } & 
\multicolumn{1}{p{2.cm}}{\hspace{0.25 cm} $T$ (days)} &
\multicolumn{1}{p{3.5cm}}{\hspace{0.5 cm} $f_{\rm GW}=\frac{2}{T}$ $(10^{-4} \rm Hz)$ \quad} &
\multicolumn{1}{p{2cm}}{ \hspace{0.35 cm}$M_{\rm WD}(M_{\odot})$ \qquad} & 
\multicolumn{1}{p{2cm}}{\hspace{0.3 cm} $M_{\rm NS}(M_{\odot})$ \qquad }  & 
\multicolumn{1}{p{2cm}}{\hspace{0.3 cm} $M_{\rm T}(M_{\odot})$ \qquad }  & 
\multicolumn{1}{p{1.5cm}}{\hspace{0.5 cm} $q$ \quad} & 
\multicolumn{1}{p{1.5cm}}{\hspace{0.15 cm} Stable? \quad} \vspace{0.05cm} \\ 
\hline 
B2303$+$46\footnotemark[1]   	& $12.34$  &  $0.0187$                   & $1.3$                 &   $ 1.34$     &   $2.64$     & $0.97$       & No                            \\  \hline
J0621$+$1002\footnotemark[2] 	& $8.32$    &  $0.0278$                   & $0.67$               & $ 1.70$      &   $2.37$      & $0.394$     & ?                             \\  \hline
J1141$-$6545\footnotemark[3]$^,$\footnotemark[4]& $0.198$  &  $1.169$                     & $1.02$               &  $ 1.27$     &   $2.29$      & $0.803$     & No     \\  \hline
B1516$+$02B\footnotemark[5]    	& $6.858$  &  $0.0337$                   & $0.13$               & $ 2.08$      &   $2.21$      & $0.0625$    & Yes                             \\  \hline
J1713$+$0747\footnotemark[1]    	& $67.8$    &  $0.0034$                   & $0.33$               &  $ 1.60$     &   $1.93$      & $0.206$     & ?                       \\  \hline
B1855$+$09\footnotemark[1]   	       & $12.3$    &  $0.0188$                   & $0.267$             &  $ 1.58$     &   $1.847$    & $0.169$     & Yes                       \\  \hline
J0437$-$4715\footnotemark[1]  	& $5.74$    &  $0.0403$                   & $0.236$             &  $ 1.58$     &   $1.816$    & $0.149$     & Yes                        \\  \hline
J1012$+$5307\footnotemark[1]  	& $0.605$  &  $0.382$                     & $0.16$                & $ 1.64$     &   $1.80$      & $0.097$     & Yes                       \\  \hline
J0751$+$1807\footnotemark[1]  	& $0.263$  &  $0.88$                       & $0.125$              & $ 1.26$     &   $1.385$    & $0.099$     & Yes                       \\  \hline\hline 
\end{tabular}
\begin{flushleft}
\footnotetext[1]{Stairs \cite{WDNS7} and references therein.}  
\footnotetext[2]{Nice et al. \cite{WDNS2}.}   
\footnotetext[3]{Bhat et al. \cite{WDNS4}.}  
\footnotetext[4]{Bailes et al. \cite{WDNS5}.}   
\footnotetext[5]{Freire et al. \cite{WDNS6}.}   
\end{flushleft}
\label{table1}
\end{table*}

\begin{table*} [t]
\caption{WDNS binaries which do not have well-determined masses. From left to right the columns give the name of the object, the orbital period, the corresponding quadrupole GW frequency, the WD mass, the NS mass, the total mass, the mass ratio $q=M_{\rm WD}/M_{\rm NS}$ and the stability of mass transfer after its onset.}
\begin{tabular}{lccccccc}\hline\hline
\multicolumn{1}{p{2.15 cm}}{\hspace{0.3 cm} PSR } & 
\multicolumn{1}{p{2.cm}}{\hspace{0.25 cm} $T$ (days)} &
\multicolumn{1}{p{3.5cm}}{\hspace{0.5 cm} $f_{\rm GW}=\frac{2}{T}$ $(10^{-4} \rm Hz)$ \quad} &
\multicolumn{1}{p{2cm}}{ \hspace{0.35 cm}$M_{\rm WD}(M_{\odot})$ \qquad} & 
\multicolumn{1}{p{2cm}}{\hspace{0.3 cm} $M_{\rm NS}(M_{\odot})$ \qquad }  & 
\multicolumn{1}{p{2cm}}{\hspace{0.3 cm} $M_{\rm T}(M_{\odot})$ \qquad }  & 
\multicolumn{1}{p{1.5cm}}{\hspace{0.5 cm} $q$ \quad} & 
\multicolumn{1}{p{1.5cm}}{\hspace{0.15 cm} Stable? \quad} \vspace{0.05cm} \\
\hline 
J1435$-$60\footnotemark[1]   	& $1.355$  &  $0.1708$                   & $1.10$                 & $ 1.40$      &   $2.50$      & $0.785$         & No                             \\  \hline
J1157$-$5114\footnotemark[2]   	& $3.507$  &  $0.066$                     & $1.14$                 &  $ 1.35$     &   $2.49$      & $0.844$         & No                             \\ \hline
J1453$-$58\footnotemark[1]     	& $12.42$  &  $0.0186$                   & $1.07$                 &   $ 1.40$     &   $2.47$     & $0.764$       & No                            \\  \hline
J1022$+$1001\footnotemark[1]    & $7.805$  &  $0.0296$                   & $0.872$               & $ 1.40$    &   $2.272$     & $0.623$       & No                             \\  \hline
B0655$+$64\footnotemark[1]    	& $1.029$   &  $0.2948$                   & $0.814$               &  $ 1.40$     &   $2.214$      & $0.581$         & No                       \\  \hline
J2145$-$0750\footnotemark[1]   	& $6.839$    &  $0.0338$                   & $0.515$               &  $1.40$     &   $1.915$    & $0.368$         & ?                       \\  \hline
J1757$-$5322\footnotemark[2]  	& $0.453$    &  $0.511$                   & $0.55$              &  $1.35$     &   $1.90$    & $0.407$         & ?                        \\  \hline
J1603$-$7202\footnotemark[1]  	& $6.309$    &  $0.0367$                   & $0.346$              &  $ 1.40$     &   $1.746$    & $0.247$         & ?                        \\  \hline
J1810$-$2005\footnotemark[1]    & $15.01$    &  $0.0154$                   & $0.34$              &  $ 1.40$     &   $1.74$    & $0.243$         & ?                        \\  \hline
J1904$+$04\footnotemark[1]  	& $15.75$  &  $0.0147$                     & $0.27$                & $ 1.40$     &   $1.67$      & $0.193$         & ?                       \\  \hline
J1232$-$6501\footnotemark[1]    & $3.507$  &  $0.066$                       & $0.175$              & $ 1.40$     &   $1.575$    & $0.125$         & Yes                       \\  \hline\hline 
\end{tabular}
\begin{flushleft}
\footnotetext[1]{Tauris et al. \cite{WDNS1} and references therein.}  
\footnotetext[2]{Edwards and Bailes \cite{WDNS3}.}   
\end{flushleft}
\label{table2}
\end{table*}

Population synthesis calculations by Nelemans et al.\ \cite{Nelemans01} show that there are
 about $2.2\times 10^{6}$ WDNS binaries in our Galaxy, and that they have a merger rate of $1.4\times 10^{-4} \rm yr^{-1}$. Furthermore, Nelemans et al.\ find that after a year of integration, LISA should be able to detect $128$ WDNS binaries and, after considering the contribution of the double WD background GW noise, resolve $38$ of these. On the other hand, calculations by Cooray \cite{Cooray2004}, give much more conservative numbers of resolved WDNS binaries. In particular, Cooray finds that the number of LISA-resolved WDNS binaries ranges between $1$--$10$, using a WDNS merger rate between $10^{-6} \rm yr^{-1}$--$10^{-5} \rm yr^{-1}$. Cooray's upper limit was based on merger rates calculated by Kim et al.\ \cite{Kim2004}. 

In this work we focus on WDNS binaries in close binary separations, and examine the termination point of quasiequilibrium sequences describing such binaries.  Several different astrophysical scenarios can result in such a termination point for binaries in general, including {\em direct contact} of the binary components, {\em Roche lobe overflow} by one of the two companions, or the binary reaching an {\em innermost stable circular orbit} (ISCO).  As we will find, quasiequilibrium sequences of WDNS terminate when the WD fills its Roche lobe, resulting in mass transfer from the WD onto the NS across  the inner Lagrange point.  This mass transfer can either be stable (SMT) or unstable (UMT).  We also refer to the latter scenario as the tidal disruption of the WD by the NS.  

To determine which of these two outcomes is likely for a given system, we will follow the approach of
Verbunt and Rappaport \cite{Verbunt88} and Faber et al. \cite{Faber}. As indicated in our Tables~\ref{table1} and~\ref{table2}, we shall find that among the observed WDNS binaries, some will undergo SMT, while others will undergo tidal disruption (i.e.~UMT). Note that mass transfer stability has also been studied in \cite{Hut,Marsh}.

In the case of SMT, the orbital evolution of the binary occurs on a secular
 timescale determined by the emission of gravitational radiation. 
Therefore, the quasistationary conservative treatment of Clark and Eardley \cite{Clark77} and Faber et al. \cite{Faber} is adequate to
follow the evolution during this secular phase.  
Note that a quasistationary treatment to follow the orbital evolution of binary systems has been employed by several authors in the past. For example,
Rappaport et al.~\cite{Rappaport82} studied compact binaries where the mass  of the secondary can be up to $\sim 1M_\odot$ and modeled as a $n=3/2$
polytrope. Fryer et al.~\cite{Fryer99} employed a non-conservative quasistationary approach to study the evolution of white dwarf--black hole binaries,  
while Marsh et al.~\cite{Marsh} studied the evolution of double WD binary systems. Both the tidal disruption and SMT phase of double WD systems have also been studied in \cite{Benz1990, Podsiadlowski92, RasioShapiro95, Segretain1997, Guerrero2004, Yoon2007,Dan08} via SPH simulations and in \cite{Dsouza,Motl} via grid-based hydrodynamic calculations, all in Newtonian gravitation. Finally, Newtonian SPH simulations of encounters of WDs with intermediate mass BHs ($M_{\rm BH}\sim 10^3 M_\odot$) have been performed in \cite{RosswogWDBH1,RosswogWDBH2}. 
  
In the case of tidal disruption, on the other hand,
the system will evolve on a hydrodynamical (orbital) timescale. 
In this scenario the NS may plunge into the WD and spiral 
toward the center of the star liberating its gravitational potential energy as heat in the WD material. Alternatively, the NS may
be the receptacle of massive debris from the disrupted WD. 
Depending on the details of the equation of state, a cold degenerate gas can support a maximum NS rest-mass 
between $1.89 M_\odot-2.67 M_\odot$ (corresponding to  a gravitational mass between $1.65M_\odot-2.20M_\odot$) 
against catastrophic collapse if it is not rotating (the OV limit),  
about $20\%$ more mass if it is rotating uniformly  (a ``supramassive NS'', e.g.~\cite{CooST94}), and at least $50\%$ more mass if it rotates differentially (a ``hypermassive NS'') \cite{BauSS00,2004ApJ...610..941M}.  The fate of the merged WDNS then depends on the initial masses of the progenitor stars, the 
degree of mass and angular momentum loss during the WD disruption and binary merger phases, the angular momentum profile of the NS remnant and the extent to which the remnant gas is heated by shocks as it pours onto the NS and forms an extended, massive mantle.
These are issues that require a hydrodynamic simulation to resolve. 

Moreover, ascertaining whether or not the neutron star ultimately
undergoes a catastrophic collapse to a black hole (either prompt or
delayed) requires that such a simulation be performed in full general relativity.  We plan to explore some of these alternative hydrodynamical scenarios in detail in the future,  aided by simulations that employ our adaptive mesh refinement (AMR) relativistic hydrodynamics code \cite{Etienne08}.

In this paper we survey the problem in qualitative terms. In Section~\ref{Sec:Eq} we identify some of physical parameters that are likely to play a key role in the evolution of a WDNS binary system. We then 
model the secular inspiral epoch of the binary by constructing Newtonian equilibrium models of corotational WDNS binaries in close circular orbit, up to the Roche limit.  In Section~\ref{MTSvsTD} we follow the stability analysis of Verbunt and Rappaport \cite{Verbunt88} in order to determine the late-evolution of these binaries (SMT vs.~tidal disruption). 
We treat the quasistationary SMT epoch by applying the approach of Clark and Eardley \cite{Clark77} and Faber et al.~\cite{Faber} and compute the corresponding gravitational waveforms. We discuss in general terms the possible outcomes of binaries that undergo UMT and construct approximate equilibrium configurations of merged WDNS remnants. We use these models to make some predictions regarding our future fully relativistic simulations of these systems. Finally, we conclude in Section \ref{summary}, where we summarize the main findings of this work.

%
%

\section{Equilibrium configurations}
\label{Sec:Eq}

In this section we follow Faber et al.~\cite{Faber}, who studied BHNS systems, to distinguish the different timescales which are relevant for WDNS systems.   In particular, we will find that WDs are likely to rotate
synchronously with compact companions.  We then construct and describe the basic features of Newtonian equilibrium configurations of corotational WDNS binaries in circular orbit.


\subsection{Timescales}

We begin by defining the primary as the NS and the secondary as the WD. The relevant timescales are the orbital period $T$, the gravitational wave timescale of the binary $t_{\rm GW}$ and the viscous timescale $t_{\rm vis}$ of the WD. We use the dynamical timescale $t_{\rm dyn}$ of the secondary in order to rescale all other timescales. Before we proceed we define the mass ratio $q$ of the binary according to
\labeq{massratio}{
q=\frac{M_{\rm WD}}{M_{\rm NS}}.
}
We also define the compaction parameter $C$ of the WD,
\labeq{compactness}{
C=\frac{M_{\rm WD}}{R_{\rm WD}},
}
where $R_{\rm WD}$ is the WD radius.
\begin{figure*}[t]
\centering
\subfigure[{\label{fig:compaction}}]{\includegraphics[width=0.49\textwidth]{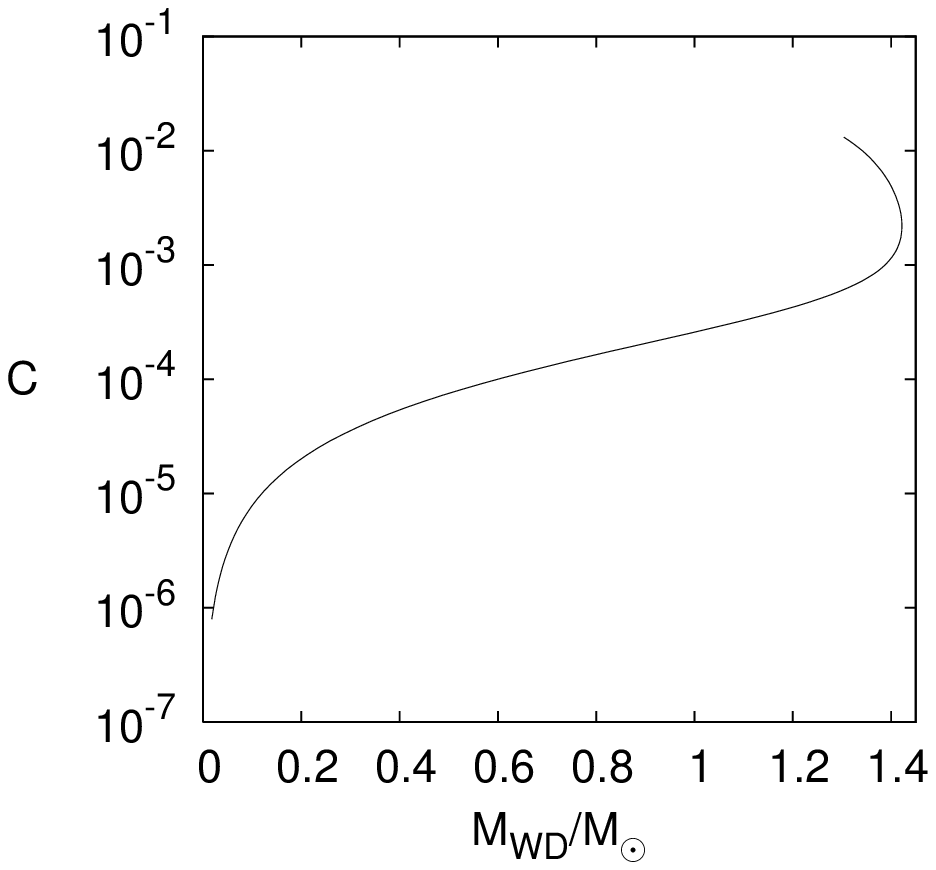}
}
\subfigure[{\label{fig:visc}}]{\includegraphics[width=0.49\textwidth]{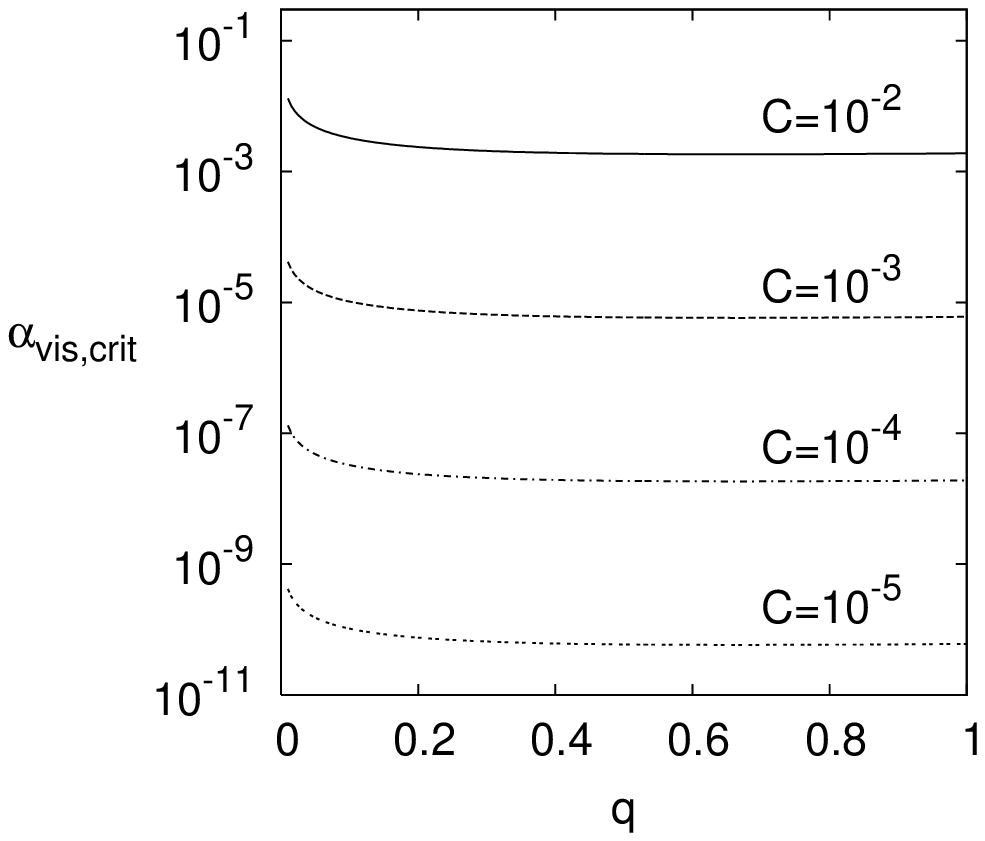}
}
\caption{{(a) Compaction $C=M_{\rm WD}/R_{\rm WD}$ of isolated WD equilibrium models plotted as a function of the WD mass. The adopted EOS is that of a cold degenerate 
ideal electron gas for $\mu_e=2$. (b) Critical turbulent viscosity parameter for a WD plotted against the mass ratio of a WDNS binary for various values of the WD compaction.
 \label{visctimescale}
 }}
\centering
\end{figure*}

By virtue of equations \eqref{massratio} and \eqref{compactness} we can define the dynamical timescale of the WD (i.e.\ the time it takes a sound wave to propagate across the star $\approx$ the free-fall timescale) as 
\labeq{tdyn}{
t_{\rm dyn} = \sqrt{\frac{R_{\rm WD}^3}{M_{\rm WD}}}=   C^{-3/2} M_{\rm WD} =  q C^{-3/2} M_{\rm NS}.
}
Note that we use geometrized units ($G=c=1$) here and throughout this paper with few exceptions where we explicitly restore cgs units.

We are interested in WDNS  systems that have reached their Roche limit. Therefore, we need an expression for  the Roche lobe ``radius" $R_{\rm r}$ of the WD. There are two approximations to the Roche lobe that have been used widely in the WD literature, namely the Paczy{\'n}ski \cite{Pacynski} and Eggleton \cite{Eggleton} formulae. We can write these in the  general form
\labeq{rocherad}{
R_{\rm r} = A f(q),
}
where $A$ is the orbital separation.  In the Paczy{\'n}ski approximation we have
\labeq{Pacynski}{
f(q) =c_0 \bigg(\frac{q}{1+q}\bigg)^{1/3}
}
with $c_0 = 0.46$, whereas in the Eggleton approximation we have
\labeq{EggleRoche}{
f(q) = \frac{c_1 q^{2/3}}{c_2 q^{2/3}+\ln(1+q^{1/3})},
}
with $c_1=0.49$ and $c_2=0.6$.

Clearly, the Paczy{\'n}ski approximation is simpler than the Eggleton approximation, but the latter is more accurate than the former \cite{Eggleton}.  For the purposes of this section it is adequate to use the Paczy{\'n}ski approximation \eqref{Pacynski}, but for our treatment of mass transfer in the following section we will adopt the Eggleton approximation \eqref{EggleRoche}.

The WD reaches its Roche limit when the Roche lobe radius becomes equal to the WD radius, $R_{\rm r}=R_{{\rm WD}}$. This occurs at orbital separation $A_{\rm R}$ given by
\labeq{aR}{
\begin{split}
A_{\rm R} =  &\ c_0^{-1} q^{-1/3}(1+q)^{1/3} C^{-1} M_{\rm WD},\\
	=  &\ c_0^{-1} q^{2/3}(1+q)^{1/3} C^{-1} M_{\rm NS}.
\end{split}
}
At this point the Keplerian orbital period is given by
\labeq{orbitalT}{
T= \frac{2 \pi}{\Omega} = 2\pi\sqrt{\frac{A_{\rm R}^3}{M_{\rm T}}}=2\pi c_0^{-3/2}q C^{-3/2} M_{\rm NS},
}
where $\Omega$ is the orbital angular frequency and $M_{\rm T}= M_{\rm WD}+M_{\rm NS}$ the total mass of the system,
so that 
\labeq{totd}{
\frac{T}{t_{\rm dyn}} = 2\pi c_0^{-3/2} \simeq 20.1
}

Treating the components of the binary as point masses and assuming the quadrupole approximation, we obtain the gravitational wave timescale
\labeq{tgrav}{
t_{\rm GW}=\frac{A}{\dot A}= \frac{5}{64}\frac{A^4}{M_{\rm NS}M_{\rm WD}M_{\rm T}},
}
where an overdot stands for a time derivative. At the critical (Roche limit) orbital separation the GW timescale is
\labeq{}{\begin{split}
t_{\rm GW}= &\ \frac{5}{64}\frac{A_{\rm R}^4}{M_{\rm NS}M_{\rm WD}M_{\rm T}}, \\
	 =	&\ \frac{5}{64}c_0^{-4}q^{5/3}(1+q)^{1/3}C^{-4} M_{\rm NS},
\end{split}
}
and hence 
\labeq{}{
\frac{t_{\rm GW}}{t_{\rm dyn}}=\frac{5}{64}c_0^{-4} q^{2/3}(1+q)^{1/3}C^{-5/2}.
}
For a realistic (massive) WD of about a solar mass, radius of $10^3\ \rm km$ (i.e. $C \simeq 10^{-3}$) and a NS companion such that $q \in [0.3,1]$ the above ratio is of order $10^7$. Thus, the GW timescale is about $10^6$ times the orbital period (see Eq. \eqref{totd}). Therefore, during the inspiral epoch (up until reaching the Roche limit) the evolution occurs on a secular timescale  and the binary system can be taken to be in a quasiequilibrium state.

We now turn to the viscous timescale $t_{\rm vis}$ of the WD. Viscosity is particularly important in determining whether the WD will be rotating synchronously in the WDNS binary or not.  According to  \cite{Bildsten92,Faber}, if the viscous timescale is short enough, tidal dissipation can synchronize the binary. In particular, this will occur if
\labeq{}{
\beta= \frac{R_{\rm WD}}{t_{\rm vis}}\geq 60 (1+q)^{5/3}q^{-2/3}C^3,
} 
or 
\labeq{}{
t_{\rm vis}\leq t_{\rm vis,crit}=\frac{1}{60} (1+q)^{-5/3}q^{5/3}C^{-4}M_{\rm NS}.
} 
According to Horedt \cite{Horedt}, turbulence is very likely to be tidally induced on the WD by the NS, when the difference between the orbital angular velocity $\Omega$ and the spin angular velocity of the WD $\omega$ is such that $|\Omega-\omega|\gtrsim 0.1 \Omega$. 
If we then assume $\alpha$-law turbulent viscosity as the primary source of viscosity, so that
\labeq{}{
\alpha_{\rm vis}=\frac{t_{\rm dyn}}{t_{\rm vis}},
}
then synchronization occurs when
\labeq{con}{
\alpha_{\rm vis}\geq \alpha_{\rm vis,crit}= 60 (1+q)^{5/3}q^{-2/3} C^{5/2}.
}
Therefore, for a given WD compaction $C$, $\alpha_{\rm vis,crit}$ increases with decreasing $q$.

The compaction of stable WDs supported by cold degenerate ideal electrons and obeying the Tolman-Oppenheimer-Volkov (TOV) equations \cite{Shapiro} lies in the interval $[10^{-6},10^{-3}]$ (see Fig.~\ref{fig:compaction}; here and in the following we assume a mean molecular weight per electron of $\mu_e = 2$). In Figure~\ref{fig:visc} we plot the critical turbulent viscosity parameter $\alpha_{\rm vis,crit}$ versus $q$ for various values of the WD compaction. For 
WD compactions  $C\lesssim5\cdot10^{-4}$, which implies $M_{\rm WD} \lesssim 1.3 M_\odot$, very small turbulent viscosity ($\alpha_{\rm vis}\simeq 10^{-6}$) is adequate to satisfy condition \eqref{con} and hence tidally lock the WD.  Plausible values of turbulent viscosity lie in the range $\alpha_{\rm vis}\in[0.001, 0.1]$ \cite{Duez}. Therefore, we conclude that the WD  component of WDNS binaries will most likely be corotational. It is interesting to contrast this result to the case of a NS in a BHNS binary, for which Faber et al.~\cite{Faber} found more likely to be irrotational.
 
Purely hydrodynamic turbulence may not be the only source of viscosity for WDs. Magnetic fields may drive MHD turbulence and thereby lock the WD tidally as well.

\subsection{Basic equilibrium equations} \label{equilibrium}

We now describe the construction of corotational equilibrium models of WDNS binary systems in order to model the early 
quasistationary inspiral epoch of the binary.

\subsubsection{Assumptions}

Here we adopt Newtonian gravitation to model the equilibrium tidal distortion of the WD. This greatly simplifies the formulation, and is reasonable due to the low compaction of WDs, $C\lesssim 10^{-3}$. The leading order relativistic corrections will be of this order, and we therefore choose to neglect them. 

To further simplify the analysis we model the NS as a point mass.  The size of the NS is negligible compared to the much larger WD, $R_{\rm NS} \ll R_{\rm WD}$, and plays little role in determining the WD profile.   Given the large compaction of the NS ($C_{\rm NS}\approx 0.1$), its deviation from sphericity will always remain very small.


\begin{figure} [h]
		\includegraphics[width=0.5\textwidth]{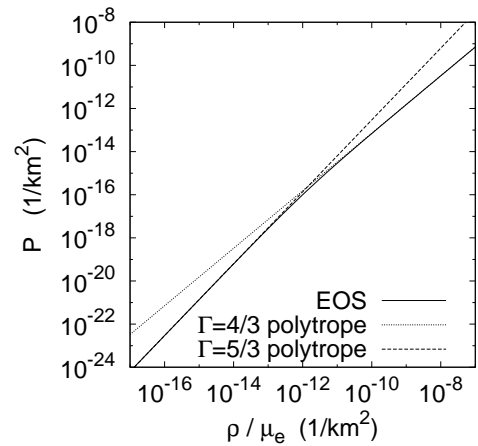}
	\caption{The exact ideal degenerate electron EOS approaches a $\Gamma=5/3$ polytrope in its low density (nonrelativistic electron) limit, and a $\Gamma=4/3$ polytrope in the high density (ultra-relativistic) regime. The transition occurs at ${\rho/\mu_e \approx 10^{-12} \rm km^{-2}}=1.347\times 10^6 \rm g\cdot cm^{-3}$. }
	\label{fig:EOS}	
\end{figure}

\subsubsection{Equation of State} \label{eq of state}

We model the WD as a synchronously rotating self-gravitating fluid supported by degenerate electron gas pressure. We assume zero temperature and ignore electrostatic corrections and effects of inverse $\beta$-decay and model the degenerate electrons as an ideal Fermi gas. We assume a mean molecular weight per electron of $\mu_e = 2$, as would be the case for a typical C-O star.  While the pressure is generated entirely by the degenerate electrons, the rest mass is contributed by the baryons.  In the extreme low and high density limits, corresponding to the nonrelativistic and the ultrarelativistic electron regime respectively, this EOS state can be approximated by a polytropic relation
	\beq \label {polytrope}
	P = K \rho ^{\Gamma},
	\eeq 
where $P$ is the pressure, $\rho$ is the (rest-mass) density, and $K$ and $\Gamma$ are constants. We denote the corresponding internal energy density $\epsilon_i$.  We graph the exact EOS, together with its two asymptotic limits, in Figure \ref{fig:EOS}. Note that restoring cgs units $1 {\rm Km}^{-2}=1.347\times 10^{18} \rm g\cdot cm^3$.

\subsubsection{Equilibrium Equations} \label{equilibrium eqns}

An equilibrium configuration of a WDNS binary is determined by the Poisson equation, 
	\beq \label{poisson}
	\nabla ^2 \phi = 4 \pi \rho,
	\eeq
 and  the integrated Euler equation, which, assuming corotation, can be written 
	\beq \label{intEuler}
	h + \phi - \frac{1}{2}\Omega^2 r_{\rm rot}^2 = C.
	\eeq
 Here $r_{\rm rot}$ is the distance from the binary's axis of rotation to a given fluid element, $C$ is a constant, and $h$ is the specific enthalpy defined as
	\beq \label{enthalpy}
	h = \int \frac{d P}{\rho}.
	\eeq
We can also write Eq. \eqref{intEuler}  as
\beq \label{intEuler2}
h = C - \phi_{\rm eff},
\eeq
where 
\beq \label{effpot}
\phi_{\rm eff} = \phi - \frac{1}{2}\Omega^2 r_{\rm rot}^2
\eeq
is the  effective potential.

The gravitational potential of the point mass NS is  
	\beq \label{phiNS}
	\phi_{\rm NS} = - \frac{M_{\rm NS}}{r_{\rm NS}},
	\eeq
where $r_{\rm NS}$ is the distance from the NS. The gravitational potential $\phi$ at any given point then can be written
	\beq \label{phi}
	\phi = \phi_{\rm WD}+\phi_{\rm NS} = \phi_{\rm WD} - \frac{M_{\rm NS}}{r_{\rm NS}},
	\eeq
where $\phi_{\rm WD}$ is the potential due to the WD.


\subsection{Numerical Technique}

We construct numerical solutions to 
Eqs. \eqref{poisson} and \eqref{intEuler2} using an algorithm based on \cite{Hac86,KomEH89a}.
Our implementation is very similar to that described in \cite{BHNS}, except that here we adopt a tabulated EOS and allow for binary companions of similar mass.

\subsubsection{Implementation of EOS}

We implemented the EOS in tabular form, using polynomial interpolation between tabulated values.
The enthalpy is computed using the tabulated values of $\rho$ and $P$ by integrating Eq. \eqref{enthalpy}. We adopted a fourth order Runge-Kutta method \cite{NR} to carry out this integration. The enthalpy $h$ is then tabulated in the final EOS table which our code reads at run time.


\subsubsection{Coordinate set up}


\begin{figure*} [t]
	\centering
		\includegraphics[width=0.9\textwidth]{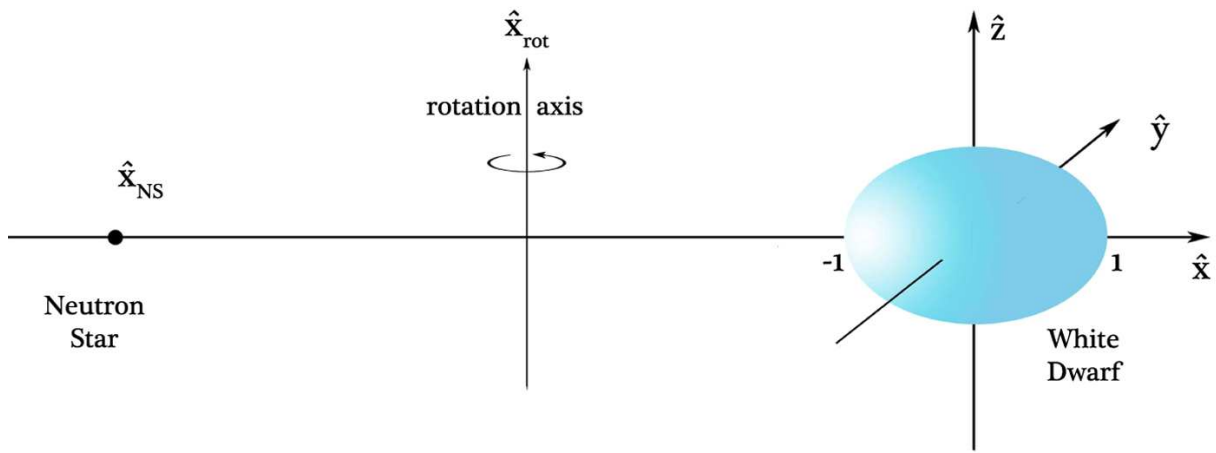}
	\caption{The coordinate system used in our calculations. We specify that the WD spans from $\hat{x}= -1$ to $\hat{x} =1$ along the $x$-axis, and rescale with each iteration. The NS is modeled as a point mass on the $x$-axis at $\hat{x}_{\rm NS}$. Our computational grid takes advantage of symmetries along the $y$ and $z$ axis, and only calculates the positive values here.}
	\label{fig:coordinatesystem}
\end{figure*}

We define Cartesian coordinates in the comoving frame of the binary following \cite{BHNS}. We assume the binary's axis of rotation to be parallel to the $z$-axis, and located at $x=x_{\rm rot}$ and $y=0$ (see Figure \ref{fig:coordinatesystem}).

It is numerically convenient to have the surface of the WD intersect the $x$-axis at fixed coordinate locations $x_A$ and $x_B$, where $x_A$ is the coordinate of the point closer to the axis of rotation. To achieve this we non-dimensionalize the coordinates using the $x$ dimension of the WD, $r_e$. For a spherical WD, $r_e$ is its radius, otherwise we calculate it as half the WD's diameter in the $x$-direction. We then define the grid coordinates as
\begin{eqnarray}\label{hat}
\hat{x} = x/r_e, \ \ \ \ \hat{y} = y/r_e, \ \ \ \ \hat{z} = z/r_e.
\end{eqnarray}
We denote grid coordinates with hats, and their physical counterparts without hats.
We choose $\hat{x}_A = -1$ and $\hat{x}_B = 1$ to define the limits of the star in the grid coordinates.

We also define
\labeq{hat2}{
\hat{\phi}_{\rm WD} = \phi_{\rm WD}/r_e^2 , \ \ \ \ \hat{\Omega}=\Omega/r_e.
}
The Laplace operator on the grid, $\hat\nabla^2$, scales as $\hat{\nabla}^2 = r_e^2 \nabla^2$. 

Given these definitions and utilizing the fact that the potential of the NS is known, we rewrite the Poisson Eq. \eqref{poisson} with respect to the computational grid as
\beq \label{code poisson}
	\hat{\nabla}^2 \hat\phi_{\rm WD} = 4 \pi \rho,
\eeq
where $\rho$ here is the rest-mass density of the WD. We also write the effective potential, Eq. \eqref{effpot}, as
\beq \label{code effpot}
\phi_{\rm eff} = \phi - \frac{1}{2}\hat{\Omega}^2\left[\left(\hat{x} - \hat{x}_{rot}\right)^2 + \hat{y}^2\right], 
\eeq
where
\beq\label{phi hat}
\phi =  r_e^2 \hat{\phi}_{\rm WD}  - \frac{M_{\rm NS}}{r_e \hat{r}_{\rm NS}}.
\eeq
Finally, we combine Eqs. \eqref{code effpot} and \eqref{phi hat} to rewrite the integrated Euler Eq. \eqref{intEuler2} as
\beq \label{code intEuler}
h = \frac{M_{\rm NS}}{r_e \hat{r}_{\rm NS}} - r_e^2 \hat{\phi}_{\rm WD} + \frac{1}{2}\hat{\Omega}^2\left[\left(\hat{x} - \hat{x}_{rot}\right)^2 + \hat{y}^2\right] + C.
\eeq
Given that the gravitational potential of the neutron star is known analytically, we may restrict the numerical grid to a region around the WD (see also \cite{BHNS}).   In our applications we impose a $1/r$ fall-off condition on $\phi_{\rm WD}$ at the outer boundaries, located at coordinate values of $\hat x = \pm 4$ and $\hat y, \hat z = 4$.  Note that for large binary separations the neutron star is outside the numerical grid.


\subsubsection{Iterative solution}

We compute equilibrium models of WDNS binaries by simultaneously solving  Eqs. \eqref{code poisson} and \eqref{code intEuler} using an algorithm that is described in detail in \cite{BHNS}.   For a given binary separation, mass ratio $q$ and central density of the WD $\rho_{\rm c}$ this iteration also determines the constants $C$, $r_e$ and the orbital angular frequency $\Omega$.

\subsubsection{Constant mass sequences} \label{Constant mass sequences}

In an effort to mimic the inspiral of the binary as it looses angular momentum to gravitational radiation, we construct sequences of equilibrium models with constant WD and NS mass and decreasing orbital separation $A$. This approach is reasonable because for large orbital separations the inspiral occurs on a secular timescale, so that at any given time, the binary can be taken to be an equilibrium configuration. 
We calculate the separation from the point source NS to the center of mass of the WD as,
\beq \label{rtot}
A = r_e (\hat{x}_{\rm NS}-\hat{x}_{\rm WD,cm}).
\eeq
The position of the NS on the coordinate grid $\hat{x}_{\rm NS}$ is stepped in sequence from large spacing (where the WD is nearly spherical) to the termination of the equilibrium sequence. The binary separation is calculated at each step from Eq. \eqref{rtot}. 

\subsection{Diagnostics and tests}

Below we summarize several physical diagnostics and tests which we use to verify the reliability of our code.

\subsubsection{Diagnostics}\label{EandJ}

We calculate the total energy according to 
\beq\label{E_eq}
E_{\rm eq} = T + W + U.
\eeq
In Eq. \eqref{E_eq} $T$ is the kinetic energy,
\begin{eqnarray} \label{T}
T &=& \frac{1}{2} \int{v^2 dm} \notag \\
	&=& \frac{1}{2} \Omega ^2 r_{\rm rot, NS}^2 M_{\rm NS} 	+ \frac{1}{2}\Omega^2 \int\limits_{\rm WD} r_{\rm rot}^2 \rho d^3 x,
\end{eqnarray}
where $r_{\rm rot, NS}$ and  $r_{\rm rot}$ are the distances of the NS and a WD fluid element from the axis of rotation respectively. The gravitational potential energy $W$ is given by,
\labeq{W}{
\begin{split}
W =& \frac{1}{2} \int{\rho \phi dV} \\
	=& \frac{1}{2} \int\limits_{\rm NS}M_{\rm NS}\delta({\bf x}-{\bf x}_{\rm NS}) \phi d^3x+\frac{1}{2} \int\limits_{\rm WD}\rho \phi d^3x\\
	=& \ \frac{1}{2} M_{\rm NS}\phi_{\rm WD}({\bf x}_{\rm NS}) 
		\ + \frac{1}{2} \int\limits_{\rm WD} \rho (\phi_{\rm NS}+\phi_{\rm WD}) d^3 x,
\end{split}
}
where $\phi_{\rm WD}({\bf x}_{\rm NS}) \approx - M_{\rm WD}/x_{\rm NS}$ is the WD potential at ${\bf x_{\rm NS}}$, the position of the NS.
 The internal energy term $U$ in Eq. \eqref{E_eq} is
\beq\label{U}
U = \int\limits_{\rm WD} \epsilon_i d^3x,
\eeq
where $\epsilon_i$ is the internal energy density of the fluid.
At infinite binary separation, which corresponds to the special case of a WD in isolation, Eq. \eqref{E_eq}  reduces to
\beq\label{E_inf}
E_{\infty} = W_{\infty} + U_{\infty},
\eeq
where $W_{\infty}$ is the potential energy of a spherical WD.
We define the system's binding energy $E_b$ as
\beq\label{E_b}
E_b = E_{\rm eq} - E_{\infty}.
\eeq
Similarly, we calculate the angular momentum of the system, $J$, as
\beq\label{J}
J = J_{\rm NS} + J_{\rm WD} =  \Omega M_{\rm NS} r^2_{\rm rot,NS} + \Omega \int\limits_{\rm WD} r^2_{\rm rot} \rho d^3 x.
\eeq


\subsubsection{A convergence test: the virial theorem}
\label{sec:virial}


\begin{figure} [t]
	\centering
		\includegraphics[width=0.5\textwidth]{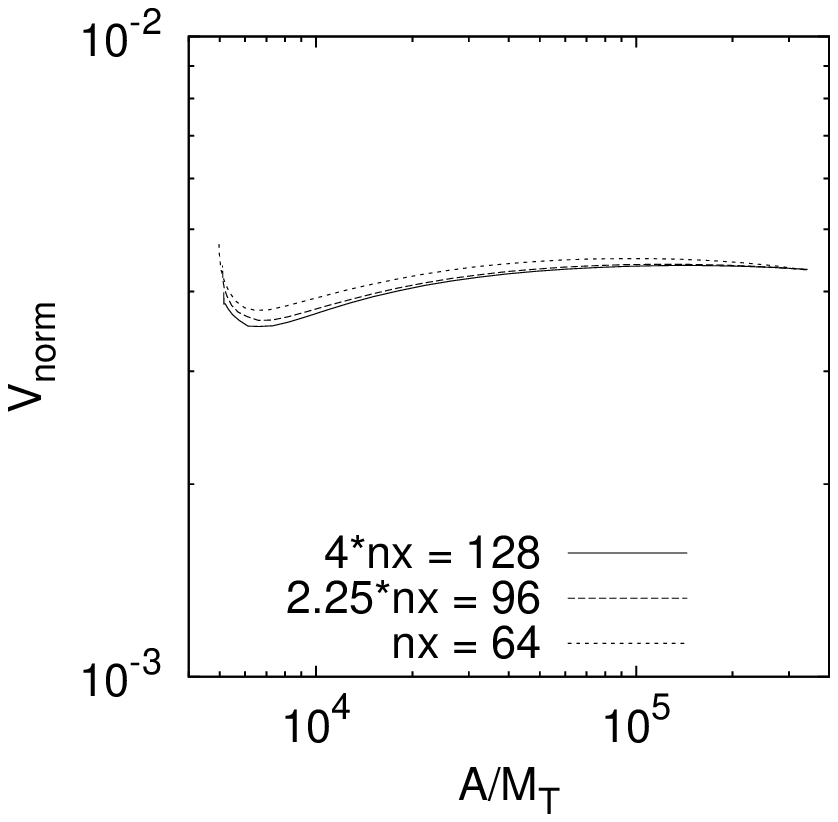}
	\caption{Normalized virial $V_{\rm norm}$ for J1141-6545 showing virial equilibrium to well within 1\% for the coarsest grid resolution, and second order convergence. Three grid resolutions are plotted, with 64x32x32, 96x48x48, and 128x64x64 grid points.}
	\label{fig:J1141_Vir}
\end{figure}

An equilibrium binary should satisfy the virial theorem. We calculate the virial expression  as an independent diagnostic test of the accuracy of our code. Following 
\cite{Shapiro}, the virial equation for our binary is
\beq \label{virial}
2T+W+3\Pi = 0,
\eeq
where $T$ and $W$ are given by Eqs. \eqref{T} and \eqref{W}. The pressure term $\Pi$ is
\beq \label{PI}
\Pi = \int\limits_{\rm WD} P d^3 x.
\eeq

A normalized virial relation that quantifies the numerical error in our calculations, is then
\beq \label{normvirial}
V_{\rm norm} = \frac {2T+W+3\Pi}{|2T|+|W|+|3\Pi |}.
\eeq
The above virial expression should converge to zero with increasing resolution and reflect the second-order accurate finite difference representation of the Poisson equation which our code employs.  In Figure \ref{fig:J1141_Vir} we show a convergence test based on the normalized virial expression for one of the constant mass sequences described in Section \ref{Constant mass sequences}, demonstrating that our code is second-order convergent.

%
%

\begin{figure*}[ht]
	\begin{tabular}{cc}
		\subfigure[]{\label{OV_Mvrho}\includegraphics[width=0.495\textwidth]{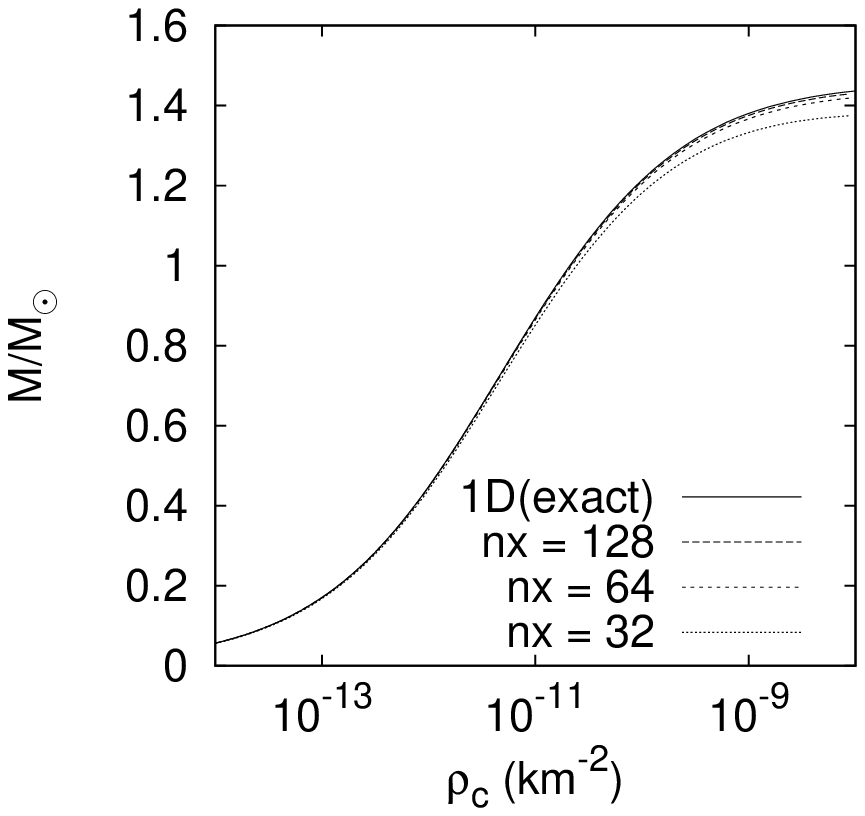}
		} &
		\subfigure[]{\label{OV_Rvrho}\includegraphics[width=0.495\textwidth]{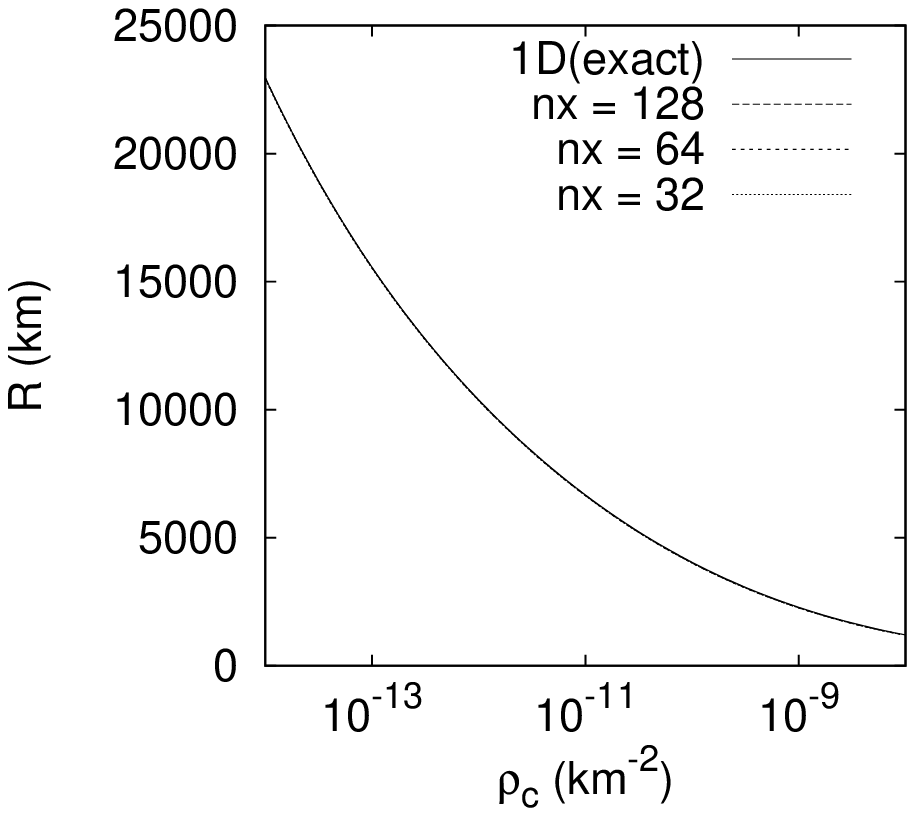}
		} \\
	\end{tabular}
	\caption{Mass $M$ (a) and radius $R$ (b) of a spherical WD versus their central density $\rho_c$, for the EOS described in Section \ref{equilibrium}. Results from our 3D code at three resolutions (corresponding to 8, 16, and 32 points across the WD radius) are plotted together with the ``exact" results from the 1D code.} 
	\label{fig:OV}
\end{figure*}


\begin{figure} [t]
    \centering
        \includegraphics[width=0.5\textwidth]{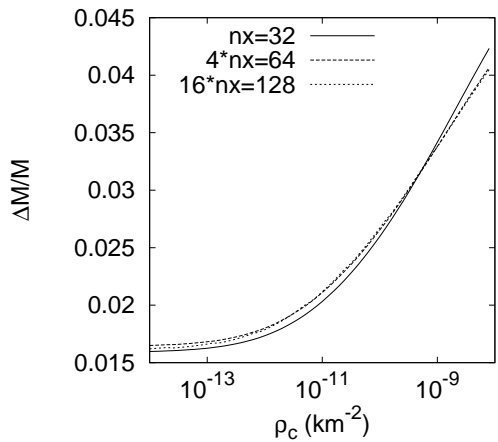}
    \caption{Relative difference between the 1D (exact) calculation of the mass for spherical WDs and the corrresponding 3D results vs. density. Here $\Delta \rm M = |\rm M-\rm M_{3\rm D}|$, where $\rm M$ is the 1D result. The plot demonstrates second order convergence of the 3D solutions towards those of the exact 1D integration.}
    \label{fig:OVconv}
\end{figure}

\subsubsection{WDs in isolation}

As another test of our code, we studied WDs in isolation and compared the results against those computed with a 1D code that takes advantage of the spherical symmetry of the WD in isolation.  To study an isolated WD with our 3D code we simply set the mass of the NS to zero.  

In Fig.~\ref{fig:OV} we show the mass and radius of the WD as a function of central density. We plot results from our 3D code, and show that they are in agreement with those obtained with our 1D code.  We label the latter, which can be obtained to machine accuracy, as ``exact". In Fig.~\ref{fig:OVconv} we also show that the masses calculated with our 3D code converge to second order toward the exact 1D results.


\begin{figure*}[htp]
  \begin{tabular} {cc}
    \subfigure[J1141-6545]{\label{fig:J1141rvOmega}\includegraphics[width=0.495\textwidth]{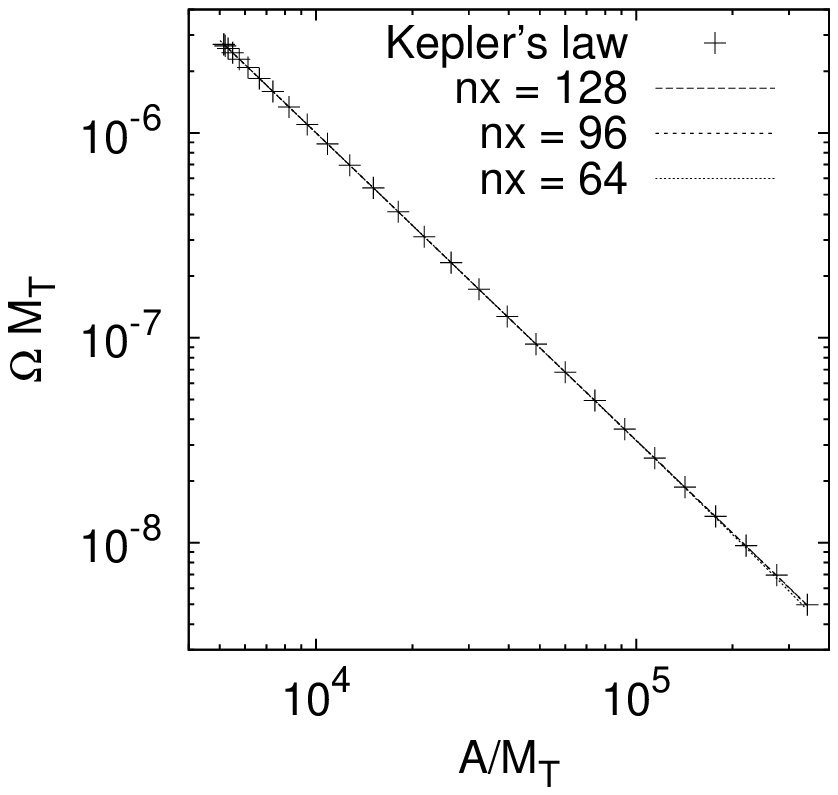}
    } &
    \subfigure[B1855+09]{\label{fig:B1855rvOmega}\includegraphics[width=0.495\textwidth]{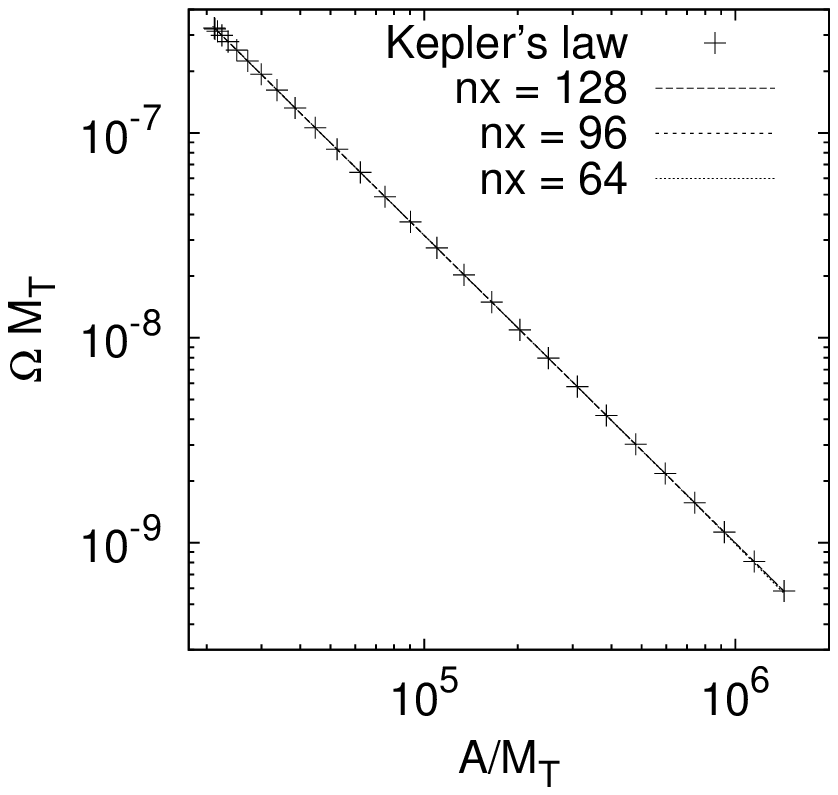}
    } \\
  \end{tabular}
  \caption{Orbital angular speed $\Omega$ plotted against binary separation $A$. $\Omega$ increases with decreasing $A$, and is closely approximated by Kepler's third law.}
  \label{fig:Omega}
\end{figure*}

\subsubsection{Orbital Angular Velocity}

In Fig.~\ref{fig:Omega} we graph the orbital angular velocity $\Omega$ versus the binary separation $A$, for two constant mass sequences representing the binaries J1141-6545 and B1855+09 (cf.~Table~\ref{table1}).   The binary J1141-6545 is a representative of the high-mass WD population, with relatively equal WD and NS masses, while B1855+09 represents the low-mass WD population (compare \cite{WDNS7}). 

Figure \ref{fig:Omega} demonstrates that the orbital angular velocity follows Kepler's third law
\beq \label{Kepler}
\Omega^2 = \frac{M_{\rm T}}{A^3}
\eeq
very closely, even when the WD is significantly distorted at small separations. This is similar to the result of \cite{RasioShapiro95}, who found that only at very small separations did the angular speed of identical binary polytropes (similar to a low mass double WD system) measurably diverge from Kepler's law.  This behavior is a consequence of the WD being centrally
condensed, so that most of its mass is concentrated close to its center.  Therefore, the orbital motion is well approximated by that for point masses, even when the envelope is significantly distorted.

\subsection{Numerical Results} \label{numresults}

With our code we produce equilibrium models of the WDNS system and quasiequilibrium sequences of constant mass that mimic the evolutionary inspiral sequences. 

\subsubsection{Roche limit}

\begin{table} [b]
\caption{The critical separation for Roche lobe overflow in units of WD volume radii $R_{\rm WD}$ for selected binaries. $A_{\rm R}^{\rm N}$ is the numerically calculated critical separation and $A_{\rm R}^{\rm P}$, $A_{\rm R}^{\rm E}$ stand for the analytically calculated $A_{\rm R}$ using Eqs. \eqref{Pacynski} and \eqref{EggleRoche} respectively.} 
\centerline{
    \begin{tabular}{lccc}
    \hline
    \hline
          Object     &\ \ \ \ \ \ \ \  $A_{\rm R}^{\rm N}$ \ \ \ \ \ \  \ \ &\  \ \ \ \ \ \  \ $A_{\rm R}^{\rm P}$ \ \  \ \ \ \ \ \  &   \ \ \ \ \ \ $A_{\rm R}^{\rm E}$\ \  \ \ \ \ \  \\
    \hline
B2303+46      &    2.66 & 2.75    & 2.66      \\
J1141-6545    &    2.79 & 2.87    & 2.82  \\    
J0621+1002    &    3.34 & 3.31    & 3.31     \\    
J1713+0747    &    4.00 & 3.91    & 3.94    \\    
B1855+09      &    4.23 & 4.13    & 4.16   \\    
J0437-4715    &    4.37 & 4.29    & 4.31   \\    
J0751+1807    &    4.94 & 4.84    & 4.85    \\    
J1012+5307    &    4.96 & 4.87    & 4.87         \\    
B1516+02B     &    5.66 & 5.58    & 5.55   \\    
    \hline
    \hline
\end{tabular} }
\label{numerical_Ar}
\end{table}



\begin{figure*} [htp] 
	\begin{tabular} {cc}
		\subfigure[J1141-6545]{	\label{fig:J1141rvBE}\includegraphics[width=0.495\textwidth]{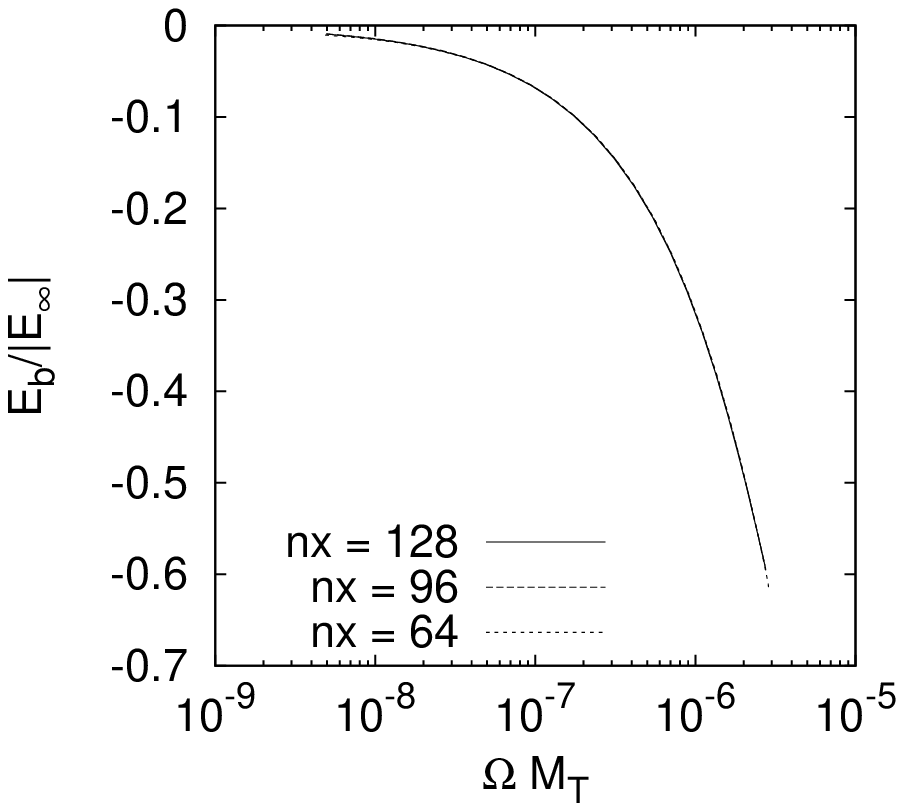}
		} &
		\subfigure[B1855+09]{	\label{fig:B1855rvBE}\includegraphics[width=0.495\textwidth]{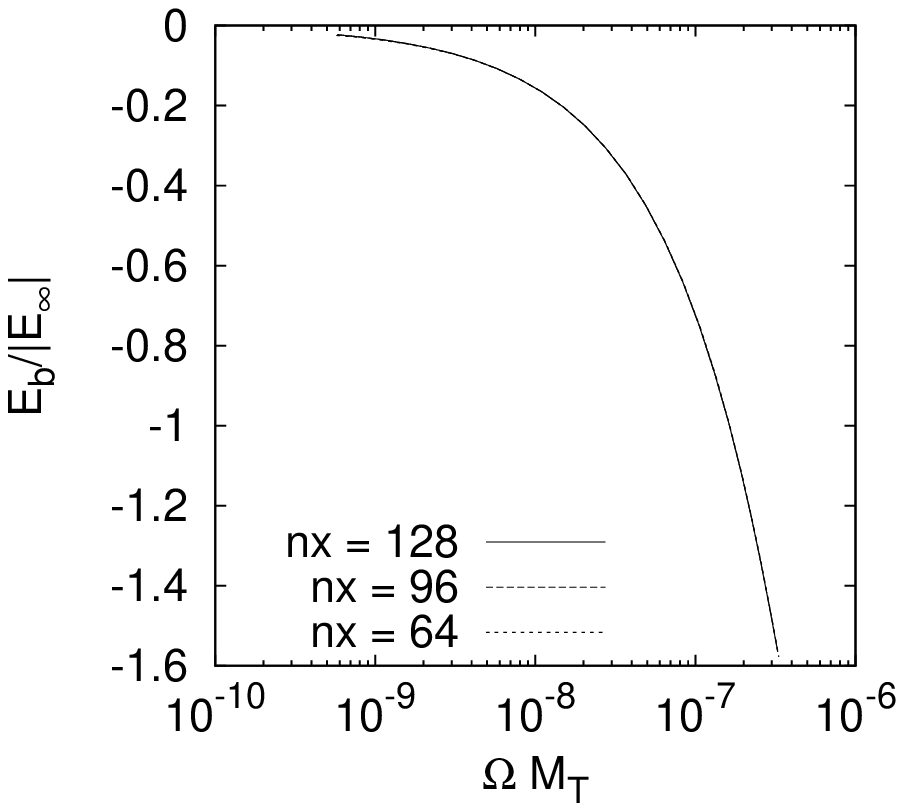}
		} \\
	\end{tabular}
	\caption{Binding energy, $E_b$ according to Eq. \eqref{E_b}, rescaled by $E_{\infty}$ as defined by Eq. \eqref{E_inf}.  The binary becomes {\em increasingly} bound with greater angular speed $\Omega$ (smaller binary separations $A$, see Fig. \ref{fig:Omega}). At very large separations, the energy of the system approaches $E_{\infty}$, as expected. At small separations we see no turning point in equilibrium energy in either case, meaning that no ISCO is encountered before reaching the Roche limit.}
	\label{fig:BEnumerical}
\end{figure*}


\begin{figure*} [htp] 
	\begin{tabular}{cc}
		\subfigure[J1141-6545]{\label{fig:J1141_rvJ}\includegraphics[width=0.495\textwidth]{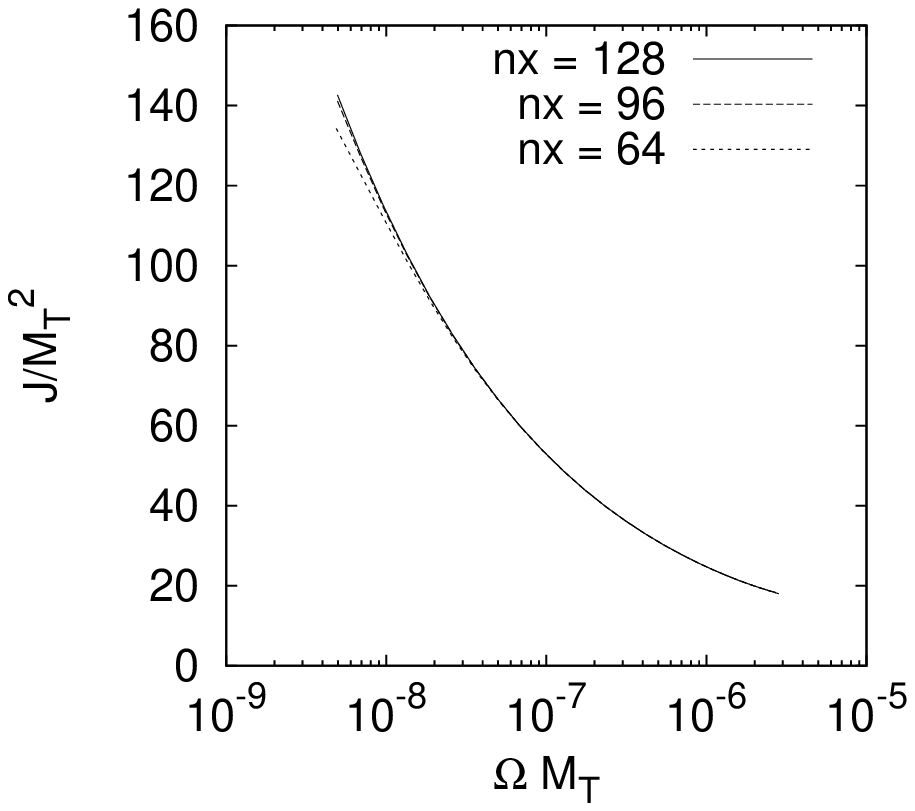}
		} &
		\subfigure[B1855+09]{\label{fig:B1855_rvJ}\includegraphics[width=0.495\textwidth]{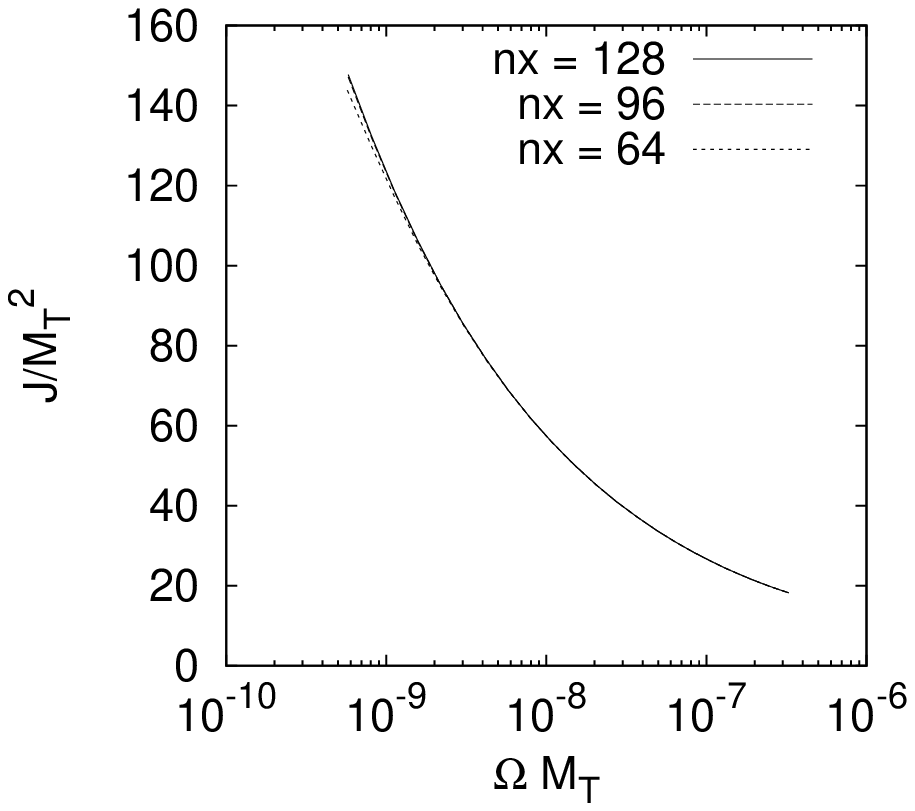}
		} \\
	\end{tabular}
	\caption{The angular momentum $J$ versus the angular speed $\Omega$. For both of these binaries, J  monotonically decreases with increasing $\Omega$ up to the termination of these plots at the Roche limit. This shows that an ISCO is {\em not} encountered before reaching the Roche limit $A_{\rm R}$.}
	\label{fig:Jnumerical}
\end{figure*}

At a certain separation our code ceases to converge and no equilibrium models can be constructed for smaller separations. At this critical separation the WD forms a cusp. This critical separation is the familiar Roche limit, $A_{\rm R}$. It is convenient to normalize the Roche limit separation with the WD volume radius, i.e., the radius of a sphere which has the same volume as the WD. At Roche limit we calculate numerically the volume $V$ of the WD and define the volume radius $R$ as 
\labeq{volradius}{
R\equiv\bigg(\frac{3V}{4\pi}\bigg)^{1/3}.
}

In Table~\ref{numerical_Ar} we tabulate the critical separation in units of the volume radius of the WD as found from the numerical model
and as predicted by the Paczy{\'n}ski Eq. \eqref{Pacynski} and Eggleton Eq.  \eqref{EggleRoche} approximations for several binaries.  The table shows that the numerical results  are in reasonable agreement with these approximations.

In Figures~\ref{fig:BEnumerical} and~\ref{fig:Jnumerical} we again focus on the binaries J1141-6545 and B1855+09 as representatives of the high and low WD mass populations.  The key result of our simulations is that for both of these classes of binaries, {\em equilibrium sequences terminate at Roche Lobe overflow}, rather than at an ISCO or at contact.  At the minimum separation plotted in the figures our sequences cease to converge, and as shown in the contour plots of the following section, the WD completely fills its Roche lobe.   We may therefore identify this binary separation with the critical binary separation $A_{\rm R}$.   The graphs of the binding energy and the angular momentum do not 
display a turning point, meaning that Roche lobe overflow occurs before the binary orbit becomes unstable at an ISCO.   Also, the minimum separation is larger than the radius of the WD and hence the sequences do not terminate at contact. 

\subsubsection{Density contours}

\begin{figure*} [t]
	\centering
		\includegraphics[width=1.0\textwidth]{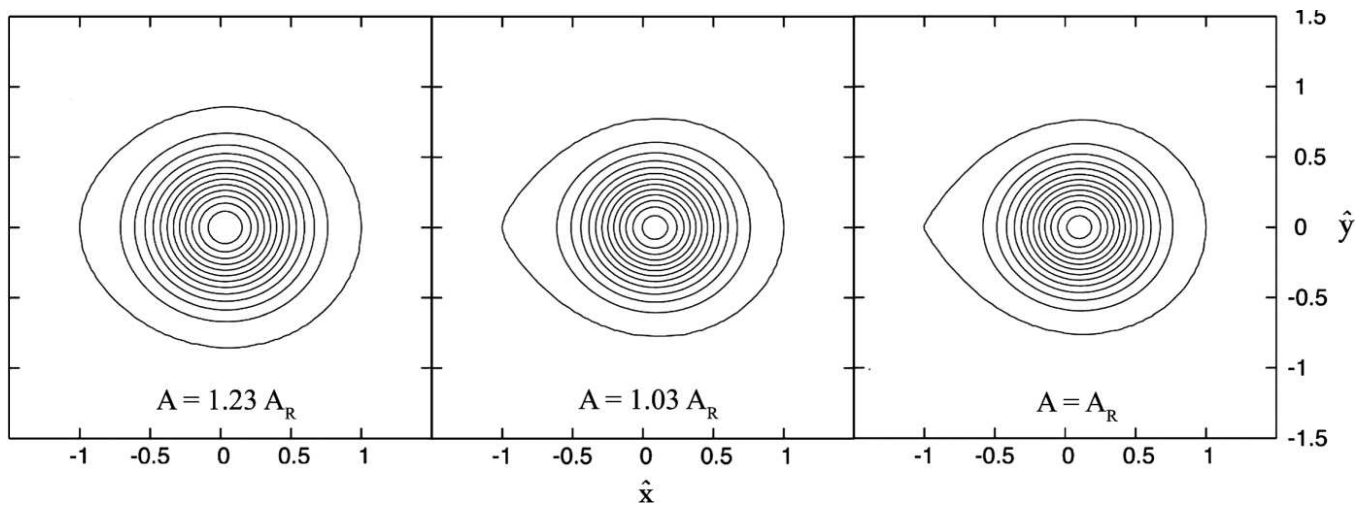}
	\caption{Contour plots of the WD in the binary B1855+09 as it approaches the Roche limit. Contours of constant density are plotted in the orbital plane. Coordinates are given in grid coordinates, and show the rescaling between iterations such that the WD always extends from $\hat{x}=1$ to $\hat{x}=-1$ along the $x$-axis. In the left panel, at $A = 1.23 A_{\rm R}$ the WD is elliptical, and contained well within its Roche lobe. In the middle panel, $A = 1.03 A_{\rm R}$, the WD is greatly distorted, and nearly fills its Roche lobe. We see a sharp cusp forming facing the NS. In the right hand panel, at $A = A_{\rm R}$, the WD exactly fills its Roche lobe, and has formed a cusp at the inner Lagrange point between the two stars. When the binary separation further decreases the WD will overflow its Roche lobe and mass will flow across this Lagrange point onto the NS.}
	\label{fig:cuspformation}
\end{figure*}

In Fig.~\ref{fig:cuspformation} we show the increasing distortion of the WD as the binary approaches the critical separation $A_{\rm R}$.   At the critical separation, the WD forms a cusp at the inner Lagrange point.  Once the binary reaches this critical separation, mass transfer from the WD onto the NS across this inner Lagrange point will occur.   

%
%

\section{Tidal disruption versus Stable Mass-Transfer}
\label{MTSvsTD}

As discussed in Section \ref{Sec:Eq}, GW-driven inspiral drives the
binary to the Roche limit, at which point mass transfer from the WD to the NS will commence.  There are at least two possible scenarios for the subsequent evolution:
\begin{itemize}
\item Mass flows on a secular timescale $\sim t_{\rm GW}$ (cf. Eq.~\ref{tgrav}) from the WD toward the NS through the inner Lagrangian point while the binary separation increases. This process is called stable mass transfer. 

\item The WD gets tidally disrupted by the neutron star, leading the system to a merger. This process comprises unstable mass transfer and its timescale is of the same order as the orbital timescale of the binary, (cf. Eq.~\ref{totd}). 
\end{itemize}
Which one of these two scenarios is likely to be realized in any particular binary is the subject of the following section.


\subsection{Stability of mass transfer}
\label{MTstab}

In this section we study the stability of mass transfer for a WDNS system near the Roche limit. Our analysis follows the approach employed by previous authors, e.g. Hut and Paczy{\'n}ski \cite{Hut} who studied low-mass semidetached binaries, Verbunt and Rappaport \cite{Verbunt}, who studied WDNS binaries, low-mass X-ray binaries and mass transfer from a (sub) giant star, and Faber et al.~\cite{Faber}, who studied BHNS binaries. Here we modify the analysis of Verbunt and Rappaport for WDNS binaries using an approximate general relativistic mass-radius relation for a cold degenerate WD, rather than the Newtonian one which was employed in \cite{Verbunt}. 

When the WD component of a WDNS binary loses mass, its radius and Roche lobe radius increase. 
The criterion for mass transfer stability  is the following: after the WD has filled its Roche lobe, instability occurs when the timescale for expansion of the WD Roche lobe radius is shorter than the timescale for expansion of the WD radius, 
\labeq{crit}{
\frac{\dot R_{\rm WD}}{R_{\rm WD}}>\frac{\dot R_{r}}{R_{r}}.
}
Before we proceed let us distinguish two different mass-exchange scenarios \cite{Verbunt,Hut}. These are the 
``no-disk" and ``with-disk" cases. 

Lubow and Shu \cite{Lubow} demonstrated that once mass transfer commences, mass flowing from the secondary does not reach the primary directly, but forms a rotating Keplerian disk around the primary.  According to \cite{Hut} the ``no-disk" scenario corresponds to the case where the disk is highly viscous and all matter lost from the WD very quickly accretes onto the NS. Assuming that the NS is not spun up, all angular momentum is then transferred very efficiently from the disk back to the orbit. In this case a massive disk onto the primary cannot form, hence the name of the scenario. This is entirely equivalent to the conservative mass-transfer case of Clark and Eardley \cite{Clark77}. The ``with-disk" scenario corresponds to the opposite case where the disk is inviscid. As a result angular momentum cannot be transfered from the disk and appreciable mass and angular momentum can be stored there forming a massive disk.

First we calculate each side of Eq. \eqref{crit} in the ``with-disk" case. Assume that mass transfer begins from the WD toward the NS at a rate $\dot M_{\rm WD}$. 
The WD radius depends only on the mass of the WD and hence
\labeq{dotRoR}{
\frac{\dot R_{\rm WD}}{R_{\rm WD}}
=\frac{d\ln R_{\rm WD}}{d\ln M_{\rm WD}}\frac{\dot M_{\rm WD}}{M_{\rm WD}}.
}

To account for the effect of the disk on the orbital motion we will make the following approximation: We assume that the combined gravitational potential and orbital angular momentum of the NS$+$disk system is the same as that of a star of mass $M=M_{\rm NS} + M_{\rm d}$, where $M_{\rm d}$ is the mass of the disk. In this approximation it is the $\rm WD$ with the $\rm NS+\rm disk$ system that is in a Keplerian orbit and not just the $\rm NS$ with the $\rm WD$. Given that the disk is gravitationally bound to the NS, this approximation is reasonable because the size of the disk is quite smaller than the binary separation. In particular, calculations by Lubow and Shu \cite{Lubow} show that for mass ratios $q<1$ the size of the disk edge is only a small fraction ($2-4\%$) of $A$.

We introduce the mass ratio $q_{\rm o}=M_{\rm WD}/M$.   
The time derivative of $q_{\rm o}$ is given by
\labeq{dotqoqii}{
\frac{\dot q_{\rm o}}{q_{\rm o}}=(1+q_{\rm o})\frac{\dot M_{\rm WD}}{M_{\rm WD}}.
} 
Using Eq.~\eqref{rocherad} the RHS of Eq.~\eqref{crit} can be written as  
\labeq{dorRroRr}{
\frac{\dot R_{r}}{R_{r}}=\frac{\dot A}{A}+\frac{d\ln f}{d\ln q_{\rm o}}\frac{\dot q_{\rm o}}{q_{\rm o}}
}
where we used $q_{\rm o}$ instead of $q$ because the WD Roche lobe now responds to the combined gravitational potential of the $\rm NS$+disk system.

We calculate $\dot A/A$ by imposing angular momentum conservation. The disk formed by mass lost from the WD, remains in a (equivalent) circular orbit whose radius $R_{\rm d}$ depends on the binary mass ratio $q$ \cite{Lubow}.  Under these assumptions the total angular momentum of the system, $J$, is given by

\labeq{Jtot}{
J=J_{\rm orb}+J_{\rm disk},
} 
where the orbital angular momentum is
\labeq{ang_mom_orb}{
J_{\rm orb}=M_{\rm T}^{3/2}A^{1/2}\frac{q_{\rm o}}{(1+q_{\rm o})^2},
}
and the disk spin angular momentum is
\labeq{disk_ang_mom}{
J_{\rm disk}=(M_{\rm NS}R_{\rm d})^{1/2} M_{\rm d}.
}
Here $M_{\rm T}=M_{\rm WD}+M$ and $R_{\rm d}=A r_h$, where $r_h$ is a function of the mass ratio $q$ given by \cite{Verbunt} 
\begin{eqnarray} \label{rh}
r_h & = &\ 0.0883-0.04858 \log q \notag\\
 &  & +0.11489 \log^2 q+0.020475  \log^3 q. 
\end{eqnarray}

Following Hut and Paczy{\'n}ski \cite{Hut} we neglect the spin angular momentum $J_{\rm s}$ of the WD for the purposes of this section. 
Our estimates show that at the Roche limit $\dot J_{\rm s}$ is typically a few
percent of either $\dot J_{\rm orb}$ or $\dot J_{\rm disk}$ because
$\dot J_{\rm s}$ scales as 
\labeq{}{
\frac{\dot J_{\rm s}}{J_{\rm orb}}= -\beta \bigg(\frac{R_{\rm WD}}{A}\bigg)^2\bigg[\frac{3}{2}\frac{\dot A}{A}-\bigg(2 \frac{d\ln R_{\rm WD}}{d\ln M_{\rm WD}}+1\bigg)\frac{\dot M_{\rm WD}}{M_{\rm WD}}\bigg],
}
where $\beta$ is a factor less than unity which accounts for the central condensation of the star and other factors of the binary mass ratio.
Note that this last equation holds true because we have assumed the binary to be corotational.
Using Eq. \eqref{Jtot}, angular momentum conservation, $\dot J =0$,  yields
\labeq{dotaoaii}{
\frac{\dot A}{A}=-2 \bigg[1-q-\left((1+q)r_h\right)^{1/2}\bigg]\frac{\dot M_{\rm WD}}{M_{\rm WD}},
}
where we have dropped terms of order $M_{\rm d}/M_{\rm NS}$, $M_{\rm d}/M_{\rm WD}$  (and hence set $q_{\rm o}=q$) because 
at the onset of mass transfer $M_{\rm d} \ll M_{\rm WD}$ and $M_{\rm d} \ll M_{\rm NS}$. Eq. \eqref{dotaoaii} is the same as that of Verbunt and Rappaport \cite{Verbunt} (note: their $q$ is the inverse of our $q$).


\begin{figure*}
\subfigure[{\label{fig:dlnRdlnMFITa}}]{ \includegraphics[width=0.48\textwidth,angle=0]{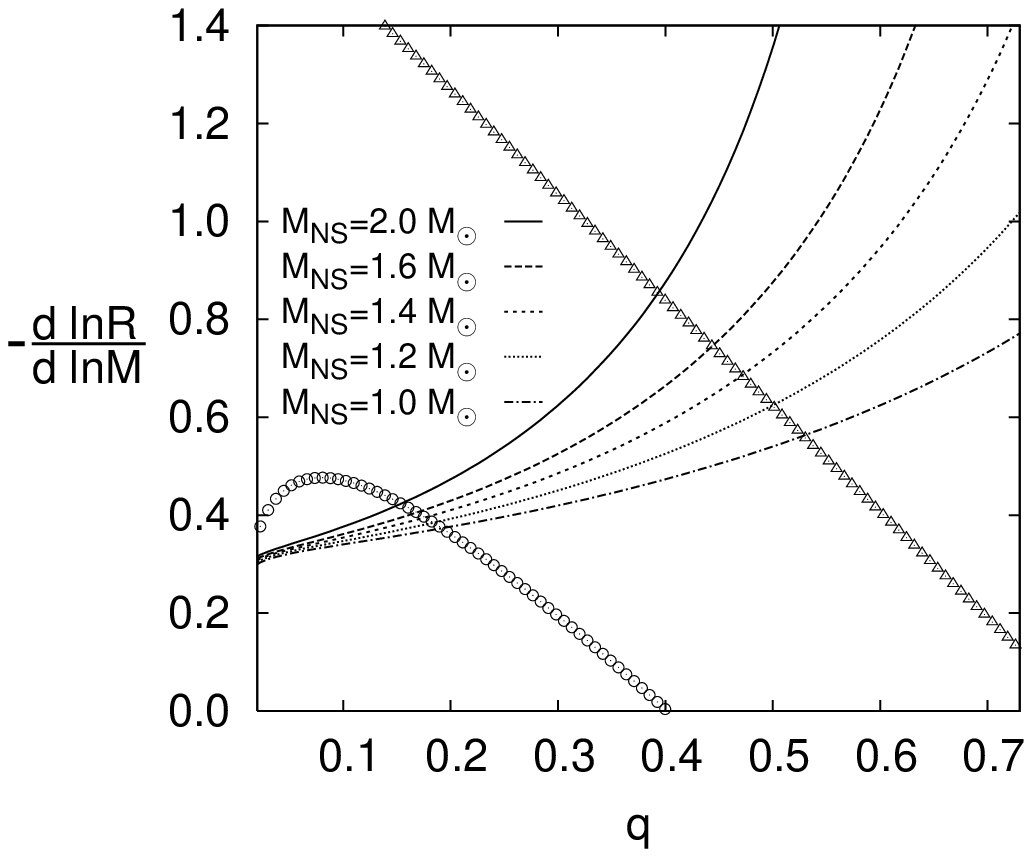}
}
\subfigure[{\label{fig:dlnRdlnMFITb}}]{ \includegraphics[width=0.48\textwidth,angle=0]{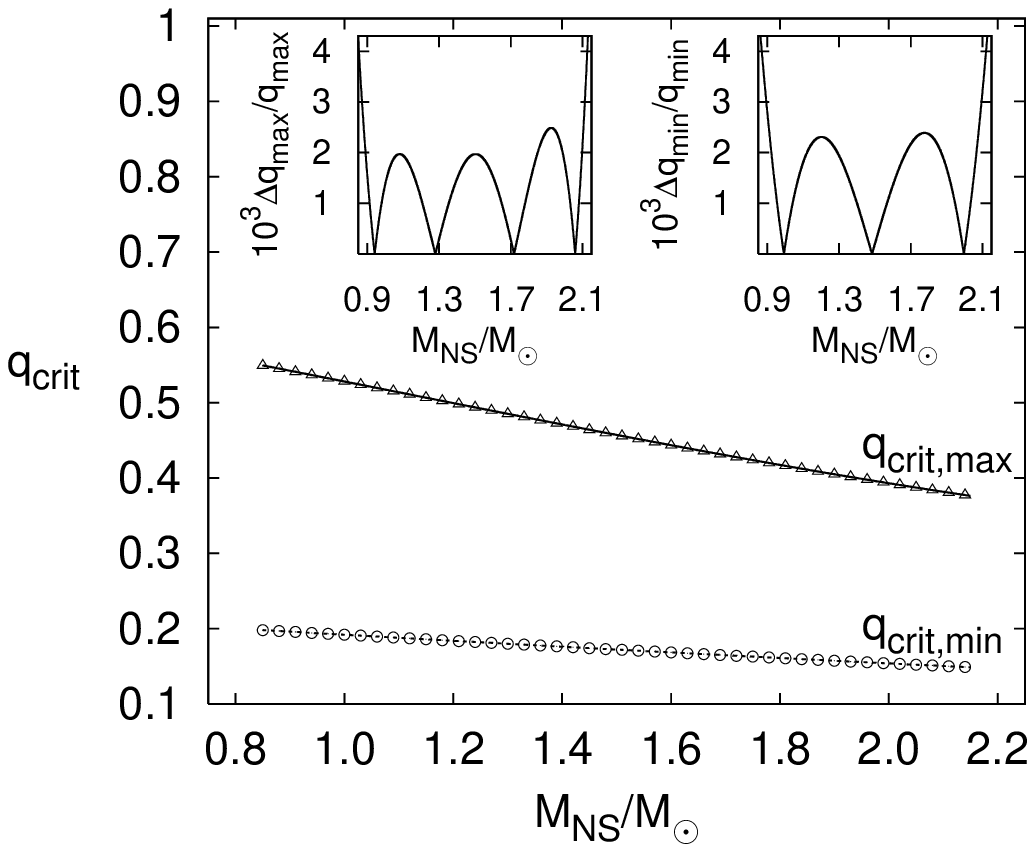}
}
\caption{(a) Graphs of $d\ln R/d\ln M$ versus $M$ relation for cold degenerate general relativistic WDs. The open triangles correspond to the RHS of Eq. \eqref{crit3} in the ``no-disk" case. The open circles correspond to the RHS of Eq.
\eqref{crit3} in the ``with-disk" case. We have used the more accurate Eggleton approximation to the Roche lobe radius for these plots.
(b) The lines represent $q_{\rm crit,max}$ and $q_{\rm crit,min}$ as given by Eqs. \eqref{qcritmax} and \eqref{qcritmin}.  The open triangles and circles represent  $q_{\rm crit,max}$ and $q_{\rm crit,min}$ obtained by the numerical solution of Eq. \eqref{crit4}. The insets show the fractional errors of the fitting formulae, where $\Delta q= |q_{\rm numerical}-q_{\rm fit}|$.  \label{fig:MTstability}
}
\end{figure*}

Combining Eqs. \eqref{crit} -- \eqref{dorRroRr} and \eqref{dotaoaii} we find that mass transfer becomes unstable if
\labeq{crit3}{
-\frac{d\ln R_{\rm WD}}{d\ln M_{\rm WD}}>2 \bigg[1-q-[(1+q)r_h]^{1/2}\bigg]-\frac{d\ln f}{d\ln q}(1+q).
}

Note that although our derivation so far corresponds to the ``with-disk" scenario,  Eq. \eqref{crit3} reduces to the ``no-disk" scenario, if we set $r_h=0$.

To complete the derivation we need to find the logarithmic derivative of the WD radius, $d\ln R_{\rm WD}/d\ln M_{\rm WD}$. 
In \cite{Verbunt,Marsh,Kopparapu} the  mass-radius relationship for a WD was approximated via the Eggleton formula
\begin{widetext}
\labeq{Eggleton}{
\frac{R_{\rm WD}}{R_{\odot}}=c\bigg[\bigg(\frac{M_{\rm WD}}{M_{ch}}\bigg)^{-2/3}-\bigg(\frac{M_{\rm WD}}{M_{ch}}\bigg)^{2/3}\bigg]^{1/2}
                   \bigg[1+d\bigg(\frac{M_{\rm WD}}{M_{p}}\bigg)^{-2/3}+\bigg(\frac{M_{\rm WD}}{M_{p}}\bigg)^{-1}\bigg]^{-2/3},
}
\end{widetext}
where $c=0.0114$, $d=3.5$, $M_{ch}=1.44 M_{\odot}$ is the Chandrasekhar mass and $M_{p}=0.00057 M_{\odot}$. In \cite{Verbunt} it is stated that this approximation corresponds to a
fully degenerate WD made of pure He. We have fit Eq. \eqref{Eggleton} to the mass-radius data plotted in Fig.~\ref{fig:OV}.
We find that
the WD mass-radius relationship is very accurately described by Eq. \eqref{Eggleton} for (see Appendix \ref{appA})
\labeq{EgglparamsOV}{
\begin{split}
c=0.0116, & \ \ \  d=4.6, \\
\ M_{ch}=1.456 M_{\odot},& \ \ \  M_{p}=4.8\cdot 10^{-5} M_{\odot}.
\end{split}
}

By virtue of Eqs. \eqref{crit3} and \eqref{Eggleton} we can now determine whether mass transfer will be stable or unstable. 
Following \cite{Verbunt}, in Fig.~\ref{fig:MTstability} we plot $-d\ln R_{\rm WD}/d\ln M_{\rm WD}$ versus the mass ratio of the WDNS binary for several NS masses. In addition, we show the RHS of inequality \eqref{crit3} calculated using the Eggleton approximation to the Roche lobe radius. Furthermore, we show both the ``no-disk" ($r_h=0$) and ``with-disk" ($r_h$ given by Eq. \eqref{rh}) models.
The intersection of the $-d\ln R_{\rm WD}/d\ln M_{\rm WD}$ curve with the curves corresponding to the RHS of Eq. \eqref{crit3} yields the critical mass ratio $q_{\rm crit}$. For $q>q_{\rm crit}$ we have UMT (tidal disruption).

Due to the uncertainty over which of the ``no-disk" and ``with-disk" scenarios actually describes a given binary we conclude that there is a minimum and a maximum critical mass ratio,  $q_{\rm crit,min}$ and $q_{\rm crit,max}$, respectively. When the mass ratio is such that $(q_{\rm crit,min}< q< q_{\rm crit,max})$ a more detailed study is required to determine the fate of a binary. Such a study would require precise knowledge of the disk viscosity mechanisms. This is a task beyond the scope of the current work, but a brief, qualitative discussion of this subject can be found in \cite{Verbunt}. For our purpose it suffices to say that if the mass ratio is larger than $q_{\rm crit,max}$, then UMT will set in and the WD will be tidally disrupted. If the mass ratio is smaller than $q_{\rm crit,min}$, then SMT will set in. 

For a given WDNS binary we can determine $q_{\rm crit,min}$ and $q_{\rm crit,max}$ by numerically solving the following equation

\labeq{crit4}{
\frac{d\ln R_{\rm WD}}{d\ln M_{\rm WD}}+2 \bigg[1-q-[(1+q)r_h]^{1/2}\bigg]=\frac{d\ln f}{d\ln q}(1+q).
}

We  solved Eq. \eqref{crit4} using a Newton-Raphson algorithm for 10000 binary models with NS masses in the range $[0.85, 2.15] M_\odot$. 
Using these solutions we obtained accurate fits for $q_{\rm crit,min}$ and $q_{\rm crit,max}$ as functions of the NS mass. 
We find that the following formulae provide such excellent fits.

\labeq{qcritmax}{
\begin{split}
q_{\rm crit,max}= &\ 0.658 - 0.107 M_{\rm NS} \\
            & \qquad - 0.034 M_{\rm NS}^2 + 0.011 M_{\rm NS}^3,
\end{split}
}

\labeq{qcritmin}{
q_{\rm crit,min}=  0.236 - 0.047 M_{\rm NS} - 0.003 M_{\rm NS}^2.
}\\

In Fig.~\ref{fig:dlnRdlnMFITb} we plot formulae \eqref{qcritmax} and \eqref{qcritmin}, and the numerical solutions of Eq. \eqref{crit4}. 
We find that the error with these formulae is of order $1$ part in $10^{3}$. 

With Eqs. \eqref{qcritmax} and \eqref{qcritmin} at our disposal we can now predict the possible fates of observed binaries. We have indicated this
in our tables~\ref{table1} and~\ref{table2}. When the mass ratio of a given binary falls in the range $[q_{\rm crit,min},q_{\rm crit,max}]$ we cannot predict the fate of the binary.

Finally, an interesting question to ask is what is the number of LISA-detectable galactic WDNS binaries that will undergo either SMT or tidal disruption per year. To answer the question we use the population synthesis results of Nelemans et al. \cite{Nelemans01} as follows: Figure 5 in \cite{Nelemans01} shows the distribution  of resolved WDNS binaries vs frequency and chirp mass ${\cal M}$, where
\labeq{chirp}{
{\cal M}=\frac{(M_{\rm NS}M_{\rm WD})^{3/5}}{(M_{\rm WD}+M_{\rm NS})^{1/5}}=M_{\rm NS}\frac{q^{3/5}}{(1+q)^{1/5}}.
}

We can calculate the binary mass ratio as a function of ${\cal M}$, for a given NS mass, via Eq. \eqref{chirp}. However, Nelemans et al. \cite{Nelemans01}  do not provide the NS masses of the individual LISA-detectable WDNS binaries they obtained. What they provide is the range of the NS masses which is $[1.25, 1.55]M_\odot$. They also provide the corresponding range of chirp masses, which is  $[0.4, 1.2]M_\odot$. This information is sufficient to place constraints on the number of stable or unstable 
LISA-resolved WDNS binaries.


\begin{figure}[t]
\includegraphics[width=0.495\textwidth,angle=0]{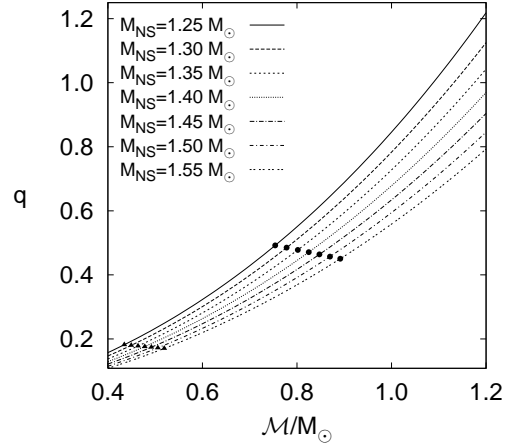}
\caption{WDNS critical chirp masses for SMT vs tidal disruption. The curves show the mass-ratio as a function of ${\cal M}$ for NS masses that cover the range of NS masses given in \cite{Nelemans01}. The filled circles and triangles represent the maximum and  minimum critical mass ratios for SMT, respectively.   \label{qdistrib}
}
\end{figure}

 We show the results of this calculation in Fig.~\ref{qdistrib}. On the $q$ vs ${\cal M}$ curves we plot the maximum and minimum critical mass ratios calculated via Eqs. \eqref{qcritmax} and \eqref{qcritmin}. 
Figure \ref{qdistrib} tells us that of the 38 LISA-resolved WDNS binaries per year, calculated in \cite{Nelemans01}, all those which have chirp mass greater than $0.89M_\odot$ will  undergo tidal disruption, whereas those which have chirp mass less than $0.43M_\odot$ will evolve into the SMT regime. The fate of those in between is uncertain.
After assessing the data given in \cite{Nelemans01} we find that after a year of integration LISA will resolve about $16$ WDNS binaries that will eventually undergo tidal disruption and 6 that will undergo SMT.  However, if the ``no-disk"  mass-transfer stability criterion is used, i.e, if we use $q_{\rm crit,max}$ as the critical value for (in)stability, we find that LISA will resolve about $20$ WDNS binaries per year that will undergo SMT.


\subsection{Stable mass transfer}
\label{stablemasstransfer}

In this section we adapt the approach of Clark and Eardley \cite{Clark77}, who studied NSNS binary systems, to follow the evolution of WDNS systems in the SMT regime. A similar approach was also used and adapted for BHNS systems in \cite{Faber}, for double WD systems in \cite{Marsh} and 
for low-mass compact binaries in \cite{Rappaport82}. 

The Clark and Eardley approach derives an evolution equation for the binary by requiring that the orbital angular momentum lost from the system be carried away by the emission of gravitational waves. However, we will also distinguish the case where angular momentum can be lost from the orbit and stored in a Lubow-Shu disk (the ``with-disk" case), in addition to the angular momentum lost in the form of gravitational radiation. Our basic assumptions are: (1) mass is conserved,
(2) mass transfer takes place across $L_1$ when the WD fills its Roche-lobe,  and (3)  the evolution takes place in quasi-equilibrium during which the gravitational radiation reaction can be modeled in the quadrupole approximation.

The sum of orbital angular momentum and the angular momentum stored in the disk is given by Eq. \eqref{Jtot}. The time derivative of $J$ is given by
\begin{widetext}
\labeq{dotJtot}{
\frac{\dot J}{J_{\rm orb}}=\bigg(1+\frac{J_{\rm disk}}{J_{\rm orb}}\bigg)\frac{\dot A}{2A} + \bigg[(1-q_{\rm o})+\frac{1}{2}\frac{d\ln r_h}{d\ln q}
\frac{J_{\rm disk}}{J_{\rm orb}}-\left(\frac{q_{\rm o}}{q}(1+q_{\rm o})r_h[q]\right)^{1/2}\bigg]\frac{\dot M_{\rm WD}}{M_{\rm WD}},
}  
\end{widetext}
where we have used Eq.~\eqref{dotqoqii}. Note that in contrast to \cite{Marsh} we retain terms of order $M_{\rm d}/M_{\rm WD}$ or $M_{\rm d}/M_{\rm WD}$. These terms are small only near the onset of mass 
transfer. Once appreciable mass is stored in the disk these terms should not be neglected.

To complete the calculation we need to link the time derivative of the orbital separation to the mass-loss rate from the WD. To do this we follow \cite{Faber} and solve Eq.~\eqref{dorRroRr} for $\dot A/A$ using the fact that $R_{\rm WD}=R_{\rm r}$ at the onset of SMT, yielding
\labeq{dotAoAevol}{
\frac{\dot A}{A}=\bigg[\frac{d\ln R_{\rm WD}}{d\ln M_{\rm WD}}-\frac{d\ln f}{d\ln q_{\rm o}}(1+q_{\rm o})\bigg]\frac{\dot M_{\rm WD}}{M_{\rm WD}}.
}

The angular momentum lost in the form of gravitational radiation $J_{\rm GW}$ is given by \cite{Shapiro}
\labeq{}{
\frac{\dot J _{\rm GW}}{J_{\rm orb}}=-\frac{32}{5}\frac{M_{\rm WD}M_{\rm NS}M_{\rm T}}{A^4}.
}
Since we assume that mass transfer takes place at the point where Roche lobe overflow occurs we have $A=A_{\rm R}=R_{\rm WD}/f(q_{\rm o})$, and hence
\labeq{dotJgwoJ}{
\frac{\dot J _{\rm GW}}{J_{\rm orb}}=-\frac{32}{5}\frac{(1+q_{\rm o})f(q_{\rm o})^4}{q_{\rm o}^2}\frac{M_{\rm WD}^3}{R_{\rm WD}^4}.
}
Angular momentum conservation implies that
\labeq{Jconserve}{
\frac{\dot J}{J_{\rm orb}}=\frac{\dot J _{\rm GW}}{J_{\rm orb}}.
}
Substituting Eqs.~\eqref{dotJtot}, \eqref{dotAoAevol} and \eqref{dotJgwoJ} into Eq.~\eqref{Jconserve}  yields the mass transfer evolution equation in the general form
\begin{widetext}
\labeq{dotM}{
\begin{split}
\dot M_{\rm WD} = & -\frac{64}{5}\bigg[\bigg(1+\frac{J_{\rm disk}}{J_{\rm orb}}\bigg)\bigg(\frac{d\ln R_{\rm WD}}{d\ln M_{\rm WD}}-\frac{d\ln f}{d\ln q_{\rm o}}(1+q_{\rm o})\bigg) \\
                 & \hspace{4cm} +2\left(1-q_{\rm o}+\frac{1}{2}\frac{d\ln r_h}{d\ln q}
\frac{J_{\rm disk}}{J_{\rm orb}}-\sqrt{\frac{q_{\rm o}}{q}(1+q_{\rm o})r_h[q]}\ \right)\bigg]^{-1}
\frac{(1+q_{\rm o})f(q_{\rm o})^4}{q_{\rm o}^2}\frac{M_{\rm WD}^4}{R_{\rm WD}^4}.
\end{split}
}
\end{widetext}

We now need to prescribe the WD
mass-radius relation and the Roche lobe radius relation. For the former we will use Eq.~\eqref{Eggleton} and for the latter we will consider the more accurate Eggleton approximation Eq.~\eqref{EggleRoche}, so that 
\labeq{}{
-\frac{d\ln f}{d\ln q}(1+q)=
\frac{q^{1/3}-q^{2/3}+q-2(1+q)\ln(1+q^{1/3})}{3[c_2q^{2/3}+\ln(1+q^{1/3})]}.
}


\begin{figure*}
\centering
\subfigure[{\label{Massevol}}]{\includegraphics[width=0.49\textwidth]{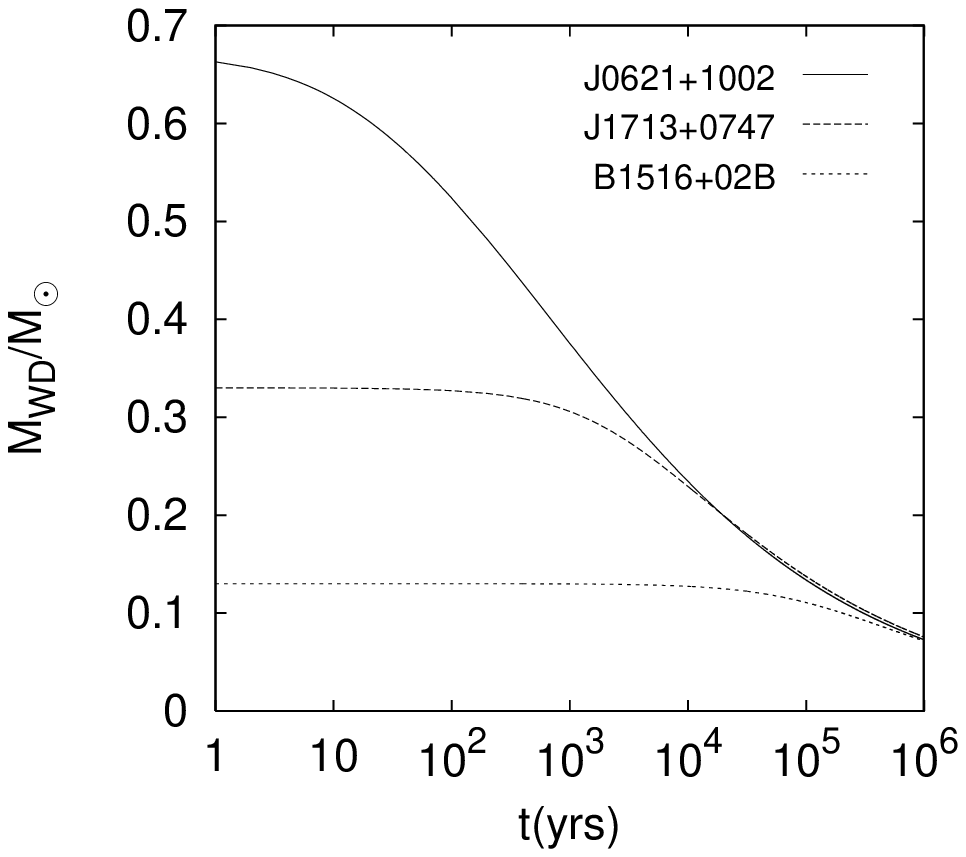}
}
\subfigure[{\label{fig:inspmst}}]{\includegraphics[width=0.49\textwidth]{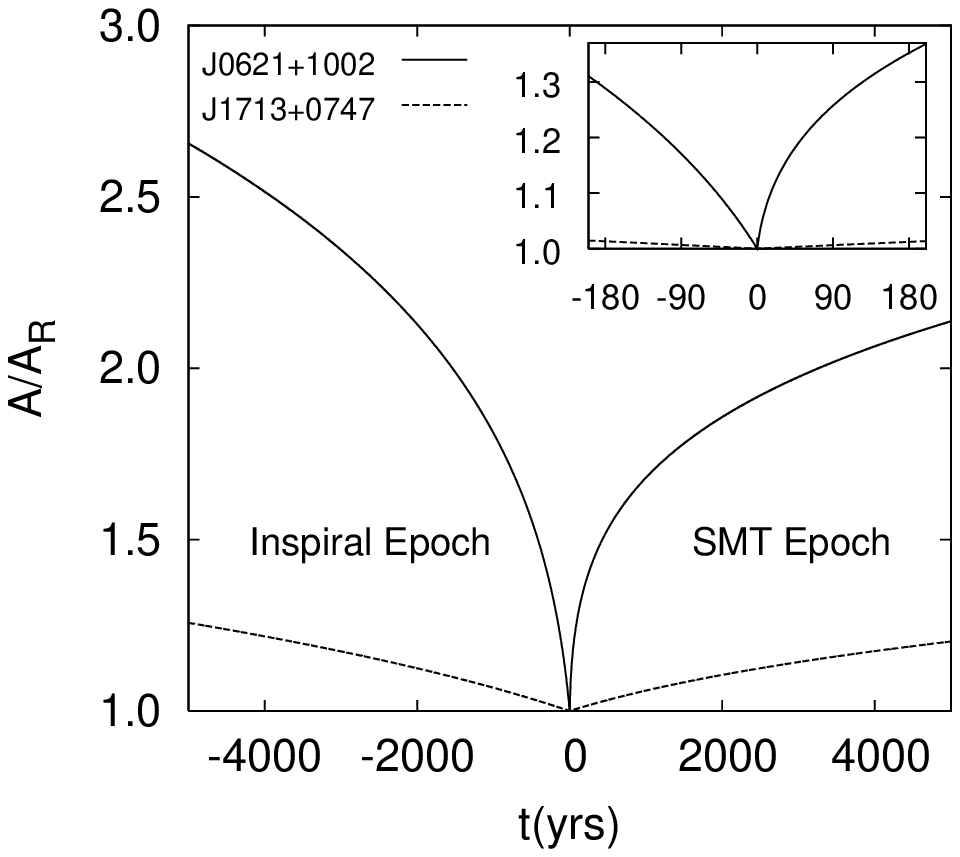}
}
\caption{{Evolution in the SMT regime assuming the ``no-disk" scenario for objects J0621$+$1002, J1713$+$0747, and B1516$+$02B. (a) Mass evolution of the WD component of the binary. (b) Evolution of the separation of the binary. The curve which corresponds to object B1516$+$02B essentially coincides with the $A/A_{\rm R}=1$ axis during the time span plotted.   
\label{MAevol}
 }}
\centering
\end{figure*}

We make the following points:
\begin{itemize}
\item Although we derived Eq. \eqref{dotM} for the ``with-disk" case, we can use it to track the evolution of the system  in the ``no-disk'' case by simply setting $r_h=0$ and $M_{\rm d}=0=J_{\rm disk}$. 

\item For SMT the initial conditions need to satisfy criterion \eqref{crit3}. Equation \eqref{crit3} arises when the term in square brackets in Eq.~\eqref{dotM} is less than zero, for $J_{\rm disk}/J_{\rm orb}\ll 1$ and $q_{\rm o}=q$; these conditions hold at the onset of mass transfer. Therefore, the closer the initial mass ratio is to the critical mass ratio for a given binary the faster this binary system will evolve.

\item Once the solution for the evolution of the WD mass is found we can find the evolution of the separation of the binary from Eq.~\eqref{dotAoAevol}.  When SMT begins the separation of the binary will increase because the term in square brackets of Eq.~\eqref{dotAoAevol} is always negative and hence the RHS of Eq.~\eqref{dotAoAevol} is always positive. 

\item When the separation is known we can also obtain the gravitational waveforms by neglecting small deviations from circular Keplerian
motion and using the quadrupole approximation. For an observer situated at $r, \theta, \phi = 0$, the ``plus", $h_{+}$, and ``cross", $h_\times$, polarizations of the gravitational waves are given by \cite{LRS94}
\labeq{hpc}{\begin{split}
r h_+ =& -2\Omega^2\mu A^2(1+\cos^2\theta)\cos2\Omega t, \\
r h_\times =& -4\Omega^2\mu A^2\cos\theta\sin2\Omega t,
\end{split}
}
where $\mu=M_{\rm WD}M_{\rm NS}/M_{\rm T}$ is the reduced mass and $\Omega=\sqrt{M_{\rm T}/A^3}$ the Keplerian angular velocity. 
\end{itemize}


\begin{center}
\begin{table}[b]
\caption{WDNS  binaries from Tables \ref{table1},~\ref{table2} that will undergo SMT and their expected GW frequencies (in units of $10^{-2}$ Hz) when Roche lobe overfill takes place and SMT sets in.}

\begin{tabular}{lc}\hline\hline
\multicolumn{1}{p{2. cm}}{\hspace{0.5 cm} PSR } & 
\multicolumn{1}{p{3.cm}}{\hspace{0.3 cm} $f_{\rm GW}$ ($10^{-2}$Hz) \quad}  \\  
\hline 
B1516$+$02B    	               & $0.57$                                \\  \hline
B1855$+$09   	             & $1.21$                            \\  \hline
J0437$-$4715  	             & $1.06$                             \\  \hline
J1012$+$5307  	               & $0.70$                          \\  \hline
J0751$+$1807  	             & $0.55$                        \\  \hline
J1232$-$6501                & $0.77$                       \\  \hline\hline 
\end{tabular}
\label{table3}
\end{table}
\end{center}

In Table~\ref{table3} we show the expected GW frequencies from binaries
of Tables \ref{table1} and \ref{table2} that will undergo SMT. All frequencies are quoted at separation $A=A_{\rm R}$. These frequencies are above the double WD background noise and fall within the LISA frequency range, which is $[10^{-4} {\rm Hz}, 1{\rm Hz}]$.

We now adopt object B1516$+$02B as representative to study whether these binaries emit detectable gravitational waves. We obtain a \emph{lower} limit on the amplitude of the waves coming from this object by assuming it is located at maximum  distance from Earth and $\theta=\pi/2$. Since these objects are all within the Galaxy we assume that the object lies at the opposite end of the Milky way with respect to our solar system. The radius of the disk of the Galaxy is about $15$ kpc and our solar system is located $8$kpc away from the center of the Galaxy. Therefore, the distance from Earth to B1516$+$02B has to be smaller than about $23$kpc. If we now use the data of Table \ref{table3} we estimate that the minimum amplitude of the expected GW at the Roche limit is $h_{\rm +,min}\simeq2\cdot 10^{-23}$. LISA's predicted strain sensitivity at $10^{-2}$ Hz is of order $h\sim 10^{-23}$.  Thus, WDNS binaries near their Roche limit could be detectable by LISA.

To determine the evolution of a given binary system we need to solve Eqs.~\eqref{dotAoAevol}  and \eqref{dotM} numerically. We adopted an adaptive step fourth-order Runge Kutta method to carry out the numerical integration. In Fig.~\ref{MAevol} we show the results of these calculations, assuming the ``no-disk" scenario and setting initial conditions corresponding to objects J0621$+$1002, J1713$+$0747, and B1516$+$02B from Tables \ref{table1} and  \ref{table2}. We choose these objects because they cover almost the entire range of observed mass ratios for SMT in the ``no-disk" case. 

In Fig.~\ref{Massevol} we show the evolution of the WD mass for these three systems. We can see that typical timescales for the WD to lose half its initial mass range between a few thousands of years (for mass ratios near the critical mass ratio) to a few millions of years (for low mass ratios). As a consequence, typical values of $\dot {\rm M}_{\rm WD}$, i.e. the WD mass-loss rate, range roughly between $10^{-4}-10^{-8}$ $\rm M_\odot/\rm yr$. If any fraction of this mass accretes onto the NS, then it would generate an X-ray photon luminosity that would accompany the GW signal from the source.


In Fig.~\ref{fig:inspmst} we show both the SMT epoch and the earlier inspiral epoch, modeled in the quadrupole approximation. The evolution equation of the binary separation during inspiral
is obtained from Eqs. \eqref{dotJtot}, \eqref{dotJgwoJ} and \eqref{Jconserve} setting $\dot M_{\rm  WD}=0$ and is given by 
\labeq{Aquadrupole}{
A^3\dot A = -\frac{64}{5}\mu M_{\rm T}^2.
}
If we integrate Eq.~\eqref{Aquadrupole} and impose the condition that $A=A_{\rm R}$ at $t=0$, we find
\labeq{Aquadrupole2}{
A=A_{\rm R} \bigg(1-4\frac{t}{t_{\rm GW}}\bigg)^{1/4}
}
Using Eq. \eqref{Aquadrupole2} we can patch the inspiral epoch to the SMT epoch. 

Fig.~\ref{MAevol} demonstrates that object J0621$+$1002 evolves on a timescale shorter than that of object J1713$+$0747 which in turn evolves on a timescale shorter than that of object B1516$+$02B. We can understand this by checking the SMT timescale, which is defined as $t_{\rm SMT}\equiv A/\dot A$, given by Eq.~\eqref{dotAoAevol}. 

Fig.~\ref{fig:inspmst} shows that in the case of object J1713$+$0747 when the binary separation is small the inspiral timescale, which is defined in Eq.~\eqref{tgrav}, is longer than the SMT timescale, while the reverse is true at large separations. To explain this behavior, we study the ratio of the inspiral timescale, $t_{\rm GW}$ to that of SMT, $t_{\rm SMT}$, at the same separation.

\begin{figure}[t]
\includegraphics[width=0.495\textwidth]{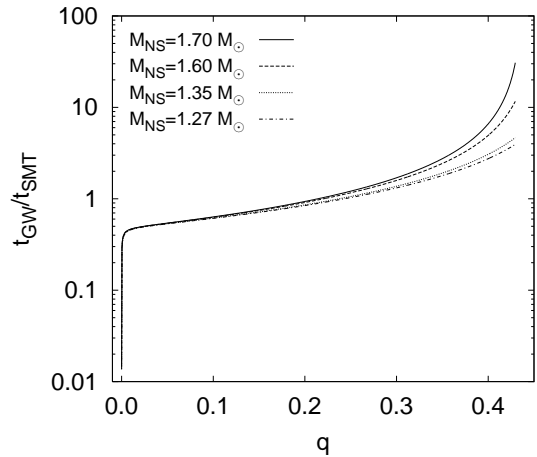}
\caption{ Inspiral timescale to SMT timescale ratio. \label{MAevol2}}
\end{figure}

In Fig.~\ref{MAevol2} we plot $t_{\rm GW}/t_{\rm SMT}$ as a function of $q$. For high $q$,  $t_{\rm SMT}\ll t_{\rm GW}$, whereas for  low $q$, $t_{\rm SMT} \gg t_{\rm GW}$. The initial conditions which correspond to  object  J1713$+$0747 yield a $q$ such that at separations close to $A_{\rm R}$, $t_{\rm GW}\simeq 2t_{\rm SMT}$. However, at large separations, after the WD has lost substantial mass and $q$ has decreased, $t_{\rm GW}$ is smaller than $t_{\rm SMT}$, which explains the behavior of the orbital evolution of J1713$+$0747. Similar reasoning explains the behavior of the orbital evolution of object B1516$+$02B. 

\begin{figure*}[t]
\subfigure[{\label{hplus}}]{\includegraphics[width=0.49\textwidth]{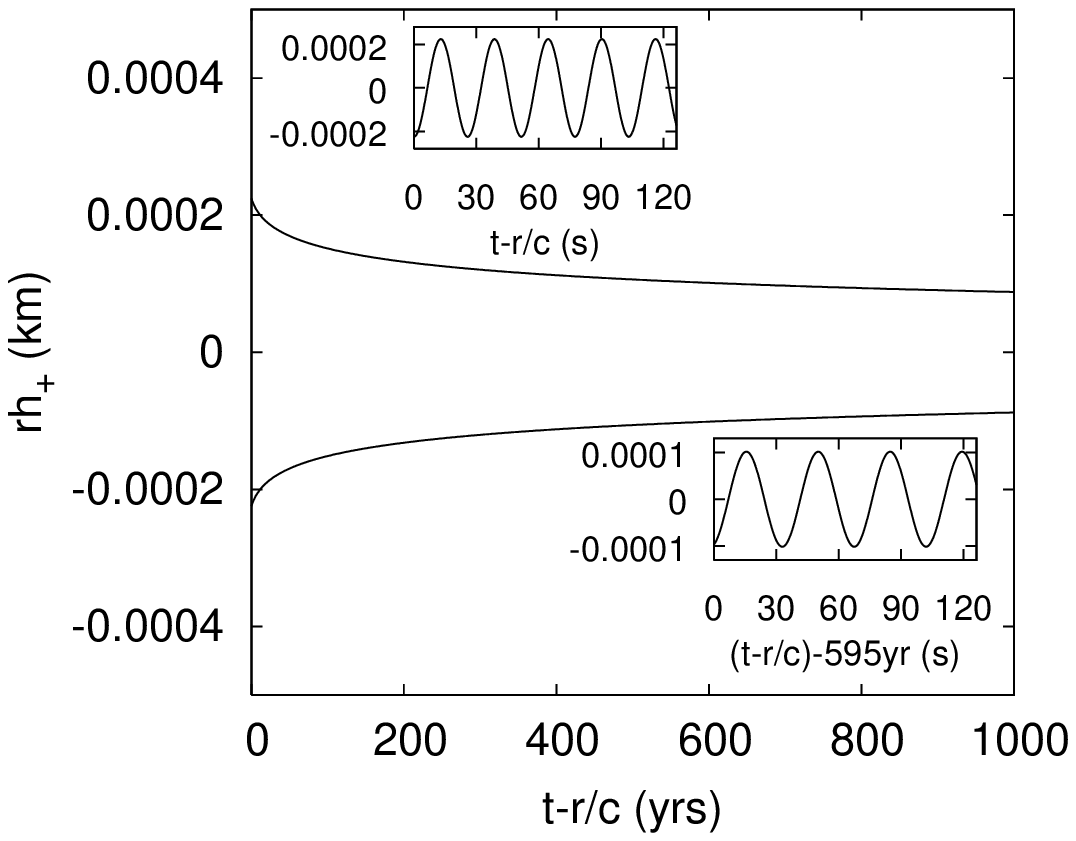}
}
\subfigure[{\label{hcross}}]{\includegraphics[width=0.49\textwidth]{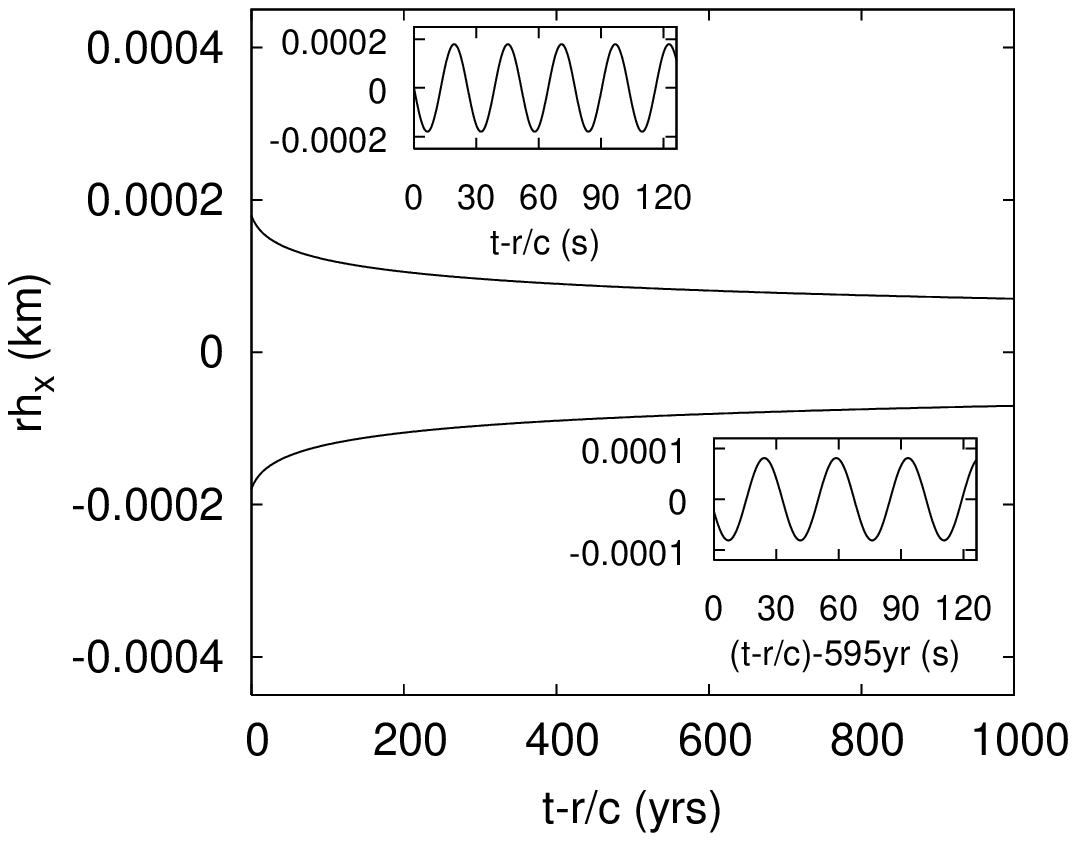}
}
\caption{{ Gravitational waveforms corresponding to the SMT evolution of object J0621$+$1002. The curves show the envelope of the wave and the insets show the  waveforms at different retarded times.  (a) The ``plus" polarization of the GW, (b) the ``cross" polarization of the GW.
 \label{Waveforms}
 }}
\end{figure*}

Once the numerical solution in the SMT regime is obtained we can calculate the gravitational waveforms from Eq.~\eqref{hpc}. In Fig.~\ref{Waveforms} we show the ``plus" and ``cross" polarizations of the gravitational waves for J0621$+$1002, setting $\theta=\pi/3$. From this figure it is clear that as a WDNS binary evolves in the SMT regime both the amplitude and the frequency of the GWs emitted decrease. The former 
results (a) because the reduced mass $\mu$ goes down, since the mass of the WD decreases with time and $q<1$, and (b) because the separation increases with time. The frequency decreases because of the increasing separation.

As mentioned, the numerical results presented so far concern the ``no-disk", $r_h=0$, case. However, the theoretical framework we presented can be applied to the ``with-disk" case as well. The results when we set $r_h\neq 0$ are qualitatively similar to those of the $r_h=0$ case and for this reason we will not present them. 


We note that there is one scenario where the quasistationary treatment of the current section is not applicable and instead a fully relativistic simulation is necessary in order to follow the evolution of a WDNS system. 
 When SMT sets in, mass accretion onto the NS may drive the NS above the TOV mass limit. 
In such a scenario the tools of numerical relativity are necessary to follow the collapse, which may be delayed, and calculate the emitted GWs.
A very interesting observational aspect of such systems is that they offer the exciting scenario that  
begins with detectable LISA signals,  
and ends with LIGO signals (from the NS collapse). Note also that object B1516$+$02B is potentially one such system because its NS has a mass of $2.08M_\odot$.


\subsection{Unstable mass transfer: Tidal disruption}
\label{tidal disruption}

As discussed in Section~\ref{MTstab}, if the mass ratio of a given WDNS binary is larger than the critical mass-ratio for mass-transfer stability, mass transfer becomes unstable. 
It is important to ask then: what are the possible outcomes for this unstable tidal disruption? This is the subject of the current section.

One possible scenario is that the  NS plunges into the WD and spirals 
toward the center of the star liberating its gravitational potential energy as heat in the WD material.
Alternatively the NS may be the receptable of massive debris from the disruption of the WD, some of which may be shock-heated 
to characteristic virial temperatures ($kT \sim Mm_{\rm p}/R\sim 10^9 \ ^{\rm o}\rm K$). In both cases  a quasistationary massive NS may be formed with a hot envelope. 
We might assign an adiabatic index of $5/3$ to the heated gas, i.e., $n = 3/2$. In this case the EOS no longer 
corresponds to that of strictly cold, extremely degenerate electrons
(for which $n$ is somewhere between $3/2$ and $3$, depending on 
the mass), but instead corresponds to semi-degenerate electrons and thermal ions. We will call this the ``hot remnant" scenario.

Notice the resemblance of such an object to another exotic class of objects, the Thorne-Zytkow Objects (TZOs) \cite{ThorneZytkow77}.
These are the end result of a merger between a NS and a red giant and might occur in the dense cores of globular clusters. The evolutionary path of a TZO involves mass accretion onto the NS which eventually exceeds the OV limit and the core collapses to form a black hole.   

Once the hot remnant cools, its fate  will depend on its mass and spin as well as the angular momentum profile.  The remnant will settle down into a stable equilibrium configuration if its total rest mass is less than that of a stable NS with the same angular momentum.  Remnants with rest masses exceeding that of  ``supramassive" NSs \cite{CooST94} ultimately collapse, unless they are supported by differential rotation (``hypermassive" NSs, \cite{BauSS00,2004ApJ...610..941M}).
Analyzing its fate requires a simulation 
in full GR, since (1) it is only an equilibrium configuration 
constructed in relativistic gravitation that exhibits a
maximum mass and (2) tracking the evolution requires a 
dynamical simulation. We intend to perform such a simulation and this will be the subject of a future work.

As a first step towards determining the properties of these remnants, we will model them as equilibrium configurations consisting of a point mass NS sitting at the center of an extended envelope formed from the WD debris. In the ``cold'' remnant scenario we shall assume mass and angular momentum (but not energy) conservation and employ 
the ellipsoidal approximation (cf. \cite{ChandrasekharEllips,Shapiro,LRS93a}) to treat rotation. Furthermore, we will assume that turbulent viscosity 
(or magnetic fields) act to drive the configuration to uniform rotation. For the ``hot'' remnant, we assume that cooling is slow so that
energy is conserved.

\subsubsection{Equilibrium configurations of a point mass surrounded by a rotating polytropic envelope}

To construct approximate Newtonian equilibrium configurations
 of rotating WDNS remnants which have a point mass NS at their center and an extended rotating envelope made of the WD debris, we closely follow the approach of
\cite{Shapiro, LRS93a} in which one defines an energy functional (the total 
energy of a configuration) and then establishes equilibria by extremizing this energy functional. The contributions to the energy are (a) the internal energy of the envelope, (b) the spin kinetic energy of the envelope, (c) the gravitational self-energy of the envelope, and (d) 
the gravitational interaction energy of the envelope with the point mass at its center. We now calculate all these components separately.

We make three main assumptions: (a) In our energy functional we adopt a polytropic density 
profile parameterized by its central value, as an approximate ``trial function'' for the 
matter in the envelope; (b) the envelope is formed entirely from material of the progenitor WD and we set its mass is equal to $M_{\rm WD}$; and (c) the remnant can be modeled as an oblate spheroid having as principal axes of its outer surface $a_1$ ($=a_2$) and $a_3$, with $a_3$ measured along the rotation axis and $a_1$  in the equatorial plane. The polytropic density profile
is assumed to be constant on self-similar spheroids.

Minimizing the energy functional with respect to central density
and eccentricity for fixed mass and angular momentum uniquely
determines the oblateness and density profile of the spheroidal 
envelope in dynamical equilibrium \cite{Shapiro, LRS93a}.

The  total internal energy $U$ of the polytropic envelope is \cite{LRS93a}
\labeq{Uinternal}{
U=k_1 K\rho_{\rm c}^{1/n} M_{\rm WD}, 
}
where $K$ is the polytropic gas constant and $n$ the polytropic index, $M_{\rm WD}$ the mass of the envelope, $\rho_{\rm c}$ its central density and 
\labeq{k1}{
k_1=\frac{n(n+1)}{5-n}\xi_1|\theta_1'|, 
}
where $\xi$ and $\theta$ are the usual Lane-Emden variables for a polytrope and $\xi_1$ and $\theta_1$ their values at the stellar surface. 

The potential self-energy $W$ of the envelope is \cite{LRS93a}
\labeq{Wself}{
W=-k_2 M_{\rm WD}^{5/3}\rho_{\rm c}^{1/3}g(e),
}
where 
\labeq{k2}{
k_2=\frac{3}{5-n}\bigg(\frac{4\pi|\theta_1'|}{\xi_1}\bigg)^{1/3}, 
}
and where 
\labeq{}{
g(e)=\frac{\sin^{-1}e}{e}(1-e^2)^{1/6}
}
is a function of the eccentricity $e$, defined as
\labeq{eccent}{
e^2=1-\bigg(\frac{a_3}{a_1}\bigg)^2.
}

The potential self-energy of the envelope can also be written as  \cite{LRS93a}
\labeq{Wself2}{
W=-\frac{3}{5-n}\frac{M_{\rm WD}^2}{R}g(\lambda).
}
Here $R$ is the volume radius of the spheroidal envelope given by 
\labeq{volumeR}{
R=(a_1^2a_3)^{1/3},
} 
so that the central density becomes
\labeq{Rrhoc}{
\rho_{\rm c} = \frac{\xi_1 M_{\rm WD}}{4\pi|\theta_1'|R^3}.
}
We have also introduced the oblateness parameter
$\lambda$ 
\labeq{lambda}{
\lambda\equiv\bigg(\frac{a_3}{a_1}\bigg)^{2/3}=(1-e^2)^{1/3},
}
in terms of which the function $g$ becomes
\labeq{goflambda}{
g(\lambda)=\lambda^{1/2}(1-\lambda^3)^{-1/2}\cos^{-1}(\lambda^{3/2}).
}

The rotational kinetic energy $T$ of the envelope is \cite{LRS93a}
\labeq{Trot}{
T = \frac{J^2}{2 I}=k_3 \lambda J^2 M_{\rm WD}^{-5/3}\rho_{\rm c}^{2/3}, 
}
where
\labeq{k3}{
k_3 = \frac{5}{4}\frac{(4\pi)^{2/3}}{\kappa_n}\bigg(\frac{|\theta_1'|}{\xi_1}\bigg)^{2/3},
}
and
\labeq{kappan}{
\kappa_n=\frac{5}{3}\frac{\int_o^{\xi_1}\theta^n\xi^4d\xi}{\xi^4|\theta_1'|}.
}
The moment of inertia $I$ is given by 
\labeq{Imom}{
I=\frac{2}{5}\kappa_n\lambda M_{\rm WD} R^2.
}

If $\phi$ is the gravitational potential due to the envelope, the gravitational interaction energy $W_i$ between the envelope and the point mass $M_{\rm NS}$, placed at the center of the envelope, is 
\labeq{Wigeneral}{\begin{split}
W_{i} & =  \int\limits_{\rm NS}\rho({\bf x})\phi({\bf x}) d^3x =   \int\limits_{\rm NS} M_{\rm NS} \delta({\bf x})\phi({\bf x}) d^3x \\
 	& =\phi(0)M_{\rm NS}, 
\end{split}
}
where $\phi(0)$ is the central gravitational potential due to $M_{\rm WD}$.

For the ellipsoidal approximation one computes first the gravitational potential energy for a spherical (nonrotating) configuration and 
then ``corrects'' this energy for rotation by multiplying by a correction function. The correction function is obtained by considering the rotating and nonrotating incompressible case (constant density everywhere). We calculate the correction function first. 

The gravitational interaction energy $W_i^{s,i}$ between an incompressible (constant density $\bar\rho$), nonrotating envelope of mass $M_{\rm WD}$ and radius $R$, and a point mass $M_{\rm NS}$
at its center is
%
\labeq{Wisphericalinc}{
W_i^{s,i} =-M_{\rm NS}\int \frac{dm}{r}=-\frac{3}{2}\frac{M_{\rm NS}M_{\rm WD}}{R}.
} 
The interaction energy $W_i^{r,i}$ between an incompressible rotating envelope of mass $M_{\rm WD}$ and volume radius $R$ and a point mass $M_{\rm NS}$
at its center is given by Eq.  \eqref{Wigeneral}, where the potential $\phi^{r,i}(0)$ at the center of the envelope is given by \cite{Shapiro}
\labeq{phirotinc}{
\phi^{r,i}(0)=-2\pi\rho a_1^2\frac{(1-e^2)^{1/2}}{e}\sin^{-1}e, 
}
where $\rho=3 M_{\rm WD}/(4\pi R^3)$. Combining Eqs. \eqref{eccent}, \eqref{volumeR}, \eqref{lambda},  \eqref{Wigeneral} and \eqref{phirotinc} we find
\labeq{Wirotinc}{
\begin{split}
W_i^{r,i} & =-\frac{3}{2}\frac{M_{\rm WD}M_{\rm NS}}{R}\frac{\sin^{-1}e}{e}(1-e^2)^{1/6} \\ 
	& =-\frac{3}{2}\frac{M_{\rm WD}M_{\rm NS}}{R}g(\lambda).
\end{split}
}
Comparison of Eqs. \eqref{Wisphericalinc} and \eqref{Wirotinc} yields the rotation correction function which is $g(\lambda)$. 

In order to find the interaction energy, $W_i$, between a rotating polytropic envelope and a point mass at its center, all we are left to do is find 
the interaction energy, $W_i^{s}$, between a spherical polytropic envelope and the same point mass located at its center, and then account for rotation by multiplying by 
$g(\lambda)$. Given Eq. \eqref{Wigeneral}, this implies that we only have to calculate the potential $\phi(0)$ of a polytrope at its center.

The gravitational potential at a point ${\bf x}$ of any arbitrary 
matter distribution $\rho({\bf x})$ is given by
\labeq{phixpolytrope}{
\phi({\bf x})=-\int \frac{\rho({\bf x'})}{|{\bf x}-{\bf x}'|}d^3x'.
}
Therefore, at ${\bf x}=0$ the potential is given by
\labeq{phi0polytrope}{
\phi(0)=-\int \frac{\rho({\bf x'})}{|{\bf x}'|}d^3x'=-\int \frac{dm}{r},
}
where $dm = 4\pi\rho r^2 dr$.
Integration of Eq. \eqref{phi0polytrope} by parts yields
\labeq{I2}{
\phi(0)=-\frac{M_{\rm WD}}{R}-\int \frac{m}{r^2}dr.
}
By using the hydrostatic equilibrium equation 
 we rewrite Eq.~\eqref{I2} as
\labeq{I3}{
\phi(0)=-\frac{M_{\rm WD}}{R}+\int \frac{1}{\rho}\frac{dP}{dr}dr.
}
Substituting Eq.~\eqref{polytrope} in Eq. ~\eqref{I3} we obtain
\labeq{I4}{\begin{split}
\phi(0)=& -\frac{M_{\rm WD}}{R}+K^{1/\Gamma}\int P^{-1/\Gamma}\frac{dP}{dr}dr \\
= & -\frac{M_{\rm WD}}{R}-(n+1)K\rho_{\rm c}^{1/n},
\end{split}
}
where $\rho_{\rm c}$ is the density at the center and 
$\Gamma=1+1/n$. Using the familiar polytropic relations
\labeq{polytropeR}{
R=\bigg[\frac{(n+1)K}{4\pi}\bigg]^{1/2}\rho_{\rm c}^{(1-n)/{2n}}\xi_1,
}
and
\labeq{polytropeM}{
M_{\rm WD}=4\pi\bigg[\frac{(n+1)K}{4\pi}\bigg]^{3/2}\rho_{\rm c}^{(3-n)/2n}\xi_1^2|\theta'_1|.
}
we obtain
\labeq{MoR}{
\frac{M_{\rm WD}}{R}=(n+1)K\rho_{\rm c}^{1/n}\xi_1|\theta'_1|.
}
Substituting Eq.~\eqref{MoR} into Eq.~\eqref{I4} then yields
\labeq{I6}{
\phi(0) = -\frac{M_{\rm WD}}{R}\bigg(1+\frac{1}{\xi_1|\theta'_1|}\bigg).
}
Finally, substituting Eq.~\eqref{I6} into Eq.~\eqref{Wigeneral} we obtain the interaction energy, $W_i^{s}$, in the spherical case
\labeq{dENS3}{
W_i^{s}=-\frac{M_{\rm NS}M_{\rm WD}}{R}\bigg(1+\frac{1}{\xi_1|\theta'_1|}\bigg).
}
As an immediate check we can insert  $\xi_1=\sqrt{6}$ and $|\theta_1'|=\sqrt{6}/3$ for the incompressible case into Eq.~\eqref{dENS3}, which yields, as expected, Eq.~\eqref{Wisphericalinc}. 

The expression for the interaction energy between the spheroidal polytropic envelope and the point mass at its center is obtained by correcting Eq.~\eqref{dENS3} for rotation, so that
\labeq{Wi}{
\begin{split}
W_i= & -\frac{M_{\rm NS}M_{\rm WD}}{R}\bigg(1+\frac{1}{\xi_1|\theta'_1|}\bigg)g(\lambda) \\
	= & -\frac{k_4}{q}M_{\rm WD}^{5/3}\rho_{\rm c}^{1/3}g(\lambda),
\end{split}
}
where we used Eq.~\eqref{Rrhoc} in the last step, and where
\labeq{k4}{
k_4= \bigg(1+\frac{1}{\xi_1|\theta'_1|}\bigg)\bigg(1+\frac{4\pi|\theta_1'|}{\xi_1}\bigg)^{1/3},
}
and, as before, $q=M_{\rm WD}/M_{\rm NS}$.

We now write the total energy functional of the point mass and spheroidal polytropic envelope as 
\labeq{TotalEnergy}{
E(M_{\rm WD}, q, J; \rho_{\rm c}, \lambda) = U + W +  T + W_i,
}
where $U$, $W$, $T$ and $W_i$ are given by Eqs. \eqref{Uinternal}, \eqref{Wself}, \eqref{Trot} and \eqref{Wi} respectively.
Given $M_{\rm WD}$, $q$ and $J$, the only parameters required to describe an equilibrium 
configuration are the central density of the envelope $\rho_{\rm c}$ and the oblateness parameter $\lambda$. The equilibrium conditions then are
\labeq{eqcond}{
\pd{E}{\rho_{\rm c}}=0 \mbox{ \ \ and \ \ } \pd{E}{\lambda}=0,
}
where the partial derivatives are taken keeping $q$, $M_{\rm WD}$ and $J$ constant.

The first equality in Eq. \eqref{eqcond} yields the virial relation
\labeq{Virial}{
\frac{3}{n}U+2 T+W_{\rm t} =0.
}
Here the total gravitational potential energy $W_{\rm t} \equiv W + W_i$ is
\labeq{Wtot}{
W_{\rm t} = -\bigg(k_2+\frac{k_4}{q}\bigg) M_{\rm WD}^{5/3}\rho_{\rm c}^{1/3}g(\lambda)= -k_5\frac{M_{\rm WD}^2}{R}g(\lambda),
}
where 
\labeq{k5}{
k_5 = \frac{3}{5-n}+\frac{1}{q}\bigg(1+\frac{1}{\xi_1|\theta_1'|}\bigg).
}

The second equality in Eq. \eqref{eqcond} yields
\labeq{ToW}{
\frac{T}{|W_{\rm t}|}=\frac{1}{2}\bigg[1+\frac{3\lambda^3}{1-\lambda^3}-\frac{3\lambda^{3/2}}{(1-\lambda^3)^{1/2}\cos^{-1}\lambda^{3/2}}\bigg],
} 
which is the usual relation between $T/|W|$ and eccentricity for a uniformly rotating spheroid (cf.~Eq.~(7.3.24) in \cite{Shapiro}) except that 
now $W$ is the \emph{total} gravitational potential energy.

By virtue of Eq. \eqref{Virial} we can obtain the mass-central density relationship
\labeq{Mrhoc}{
M_{\rm WD}=M_{\rm o}\bigg[\bigg(1+\frac{k_4}{k_2 q}\bigg)g(\lambda)\bigg(1-\frac{2T}{|W_{\rm t}|}\bigg)\bigg]^{-3/2},
}
where $M_{\rm o}$ is the mass of a nonrotating spherical polytropic star which has the same polytropic constant $K$, polytropic index $n$ and central density $\rho_{\rm c}$ as the envelope. Rewriting Eq.~\eqref{polytropeM} in terms of the constants $k_1$ and $k_2$, this mass becomes 
\labeq{Mpolytrope}{
M_{\rm o}=\bigg(\frac{3 k_1 K}{n k_2}\bigg)^{3/2}\rho_{\rm c}^{(3-n)/2n}.
} 

Using Eqs. \eqref{Rrhoc} and \eqref{Mrhoc} we can write  the equilibrium volume radius as
\labeq{Req}{
R=R_{\rm o}\bigg[\bigg(1+\frac{k_4}{k_2 q}\bigg)g(\lambda)\bigg(1-\frac{2 T}{|W_{\rm t}|}\bigg)\bigg]^{-n/(3-n)},
}
where $R_{\rm o}$ is the radius of a nonrotating spherical polytropic star with the same polytropic constant $K$, polytropic index $n$ and mass as the envelope. From Eq.~(\ref{polytropeR}) we can express $R_{\rm o}$ as 
\labeq{Rpolytrope}{
R_{\rm o}=b_n\bigg[\frac{(n+1)K}{4\pi}\bigg]^{n/(3-n)}\bigg(\frac{M_{\rm WD}}{4\pi}\bigg)^{(1-n)/(3-n)}, 
}
where 
\labeq{bn}{
b_n= \xi_1(\xi_1^2|\theta_1'|)^{-(1-n)/(3-n)}.
}

Finally, if we combine Eqs.~\eqref{TotalEnergy} and \eqref{Virial}, the equilibrium energy can be written as
\labeq{Eeq}{
E_{eq}=\frac{3-n}{n}W_{\rm t}\bigg(1-\frac{3-2n}{3-n}\frac{T}{|W_{\rm t}|}\bigg).
}

Eqs.~\eqref{Virial} and \eqref{Wtot}-\eqref{Eeq} completely determine the equilibrium configuration corresponding to a given $M_{\rm WD}$, $q$ and $J$.

\subsubsection{Initial energy and angular momentum}

We have assembled all the expressions necessary to construct approximate equilibrium configurations of  a point mass NS surrounded by an extended envelope, and will now use these expressions to model the properties of the remnants of WDNS mergers.

We first need to determine the initial energy and angular momentum of the binary system 
just before tidal disruption sets in at the Roche limit. We treat the WD as a corotating, (nearly) spherical polytrope with polytropic constant $K'$ and index $n'$ (which may be different from the $K$ and $n$ of the remnant), and model the NS as a point mass. The angular momentum of the binary is given by the following formula
\labeq{Jinit}{
J=I_{\rm cm}\Omega,
}
where $I_{\rm cm}$ is the moment of inertia of the binary calculated in the center of mass frame. We write this moment of inertia as the sum of an orbital component $I_{\rm orb}$ and the WD's spin component $I_{\rm s}$,
\labeq{Icm_sum}{
I_{cm} = I_{\rm orb} + I_{\rm s},
}
with
\labeq{Iorb}{
I_{\rm orb} = \frac{M_{\rm WD} A^2}{1+q}
}
and
\labeq{Is}{
I_{\rm s} = \frac{2}{5}\kappa_{n'}M_{\rm WD}R_{\rm WD}^2.
}
Combining the last two expressions we obtain
\labeq{Icm}{
I_{\rm cm} 
= \frac{M_{\rm WD} A^2}{1+q}\bigg[1+\frac{2}{5}\kappa_{n'}(1+q)\bigg( \frac{R_{\rm WD}}{A}\bigg)^2\bigg].
}

In Section \ref{numresults} we demonstrated that deviations from Keplerian motion are small even at the Roche limit. Thus, we approximate the angular velocity
of the binary by using Kepler's third law
\labeq{KeplerOmega}{
\Omega^2= \frac{q+1}{q}\, \frac{M_{\rm WD}}{A^3}.
} 

Using Eq.~\eqref{rocherad} we can express the critical separation $A=A_{\rm R}$ as
\labeq{rocheseparation}{
A_{\rm R}=\frac{R_{\rm WD}}{f(q)}. 
}
If we now combine Eqs.~\eqref{Jinit} through \eqref{rocheseparation} we find that the initial angular momentum becomes
\labeq{Jinit2}{
J^2=\frac{M_{\rm WD}^3 R_{\rm WD}}{q(q+1)f(q)}\bigg[1+\frac{2\kappa_{n'}(1+q)f^2(q)}{5}\bigg]^2.
}

The initial energy of the binary is given by Eq.~(7.11) in \cite{LRS93a} as the sum 
\labeq{Einit0}{
E = E_{\rm p} + T'+W_i'.
}
Here $E_{\rm p}$ is the ``intrinsic" energy of the polytrope
\labeq{E_p}{
\begin{split}
E_{\rm p} = \frac{3-n'}{3}W'\bigg[1-\bigg(\frac{3-2n'}{3-n'}\bigg)\frac{T_{\rm s}}{|W'|}\bigg],
\end{split}
}
where $W'$ is the gravitational self-energy of the WD given by
\labeq{Winit}{
W'=-\frac{3}{5-n'}\frac{M_{\rm WD}^2}{R_{\rm WD}},
}
$T_{\rm s}$ is the spin kinetic energy of the WD 
\labeq{Ts}{
T_{\rm s}=\frac{1}{2}I_{\rm s} \Omega^2,
}
$T'$ is the orbital kinetic energy, 
\labeq{T'}{
T'=\frac{(M_{\rm WD})A^2\Omega^2}{2(1+q)},
}
and $W_i'$ the interaction energy between the point mass NS and the polytropic WD
\labeq{W_i'}{
W_i'=-\frac{M_{\rm WD}M_{\rm NS}}{A},
}
where we neglect tidal interaction terms. 

Using Eq.~\eqref{KeplerOmega} we can write Eq.~\eqref{Einit0} as
\labeq{Einit}{
E =  \frac{3-n'}{3} W' -\frac{M_{\rm WD}^2}{2 q A}+\frac{3-2n'}{3}T_{\rm s},
}
By virtue of Eq.~\eqref{KeplerOmega} at $A=A_{\rm R}$ the spin kinetic energy becomes 
\labeq{Ts2}{
T_{\rm s} = \frac{\kappa_{n'}}{5}\frac{(q+1)f^3(q)}{q} \frac{M_{\rm WD}^2 }{R_{\rm WD}}
}

Combining Eqs.~\eqref{Winit}, \eqref{Einit} and \eqref{Ts2}, we can finally write the initial total energy at the Roche limit as
\labeq{Einit2}{
E=-(1-\Delta)\bigg(\frac{3-n'}{5-n'}\bigg)\frac{M_{\rm WD}^2}{R_{\rm WD}},
}
where
\labeq{Delta}{
\Delta = 1-\frac{2(3-2n')(5-n')\kappa_{n'}(q+1)f^3(q)}{15 q [2q(3-n')+f(q)(5-n')]}.
}

\subsubsection{The ``hot'' remnant}

To determine the properties of the hot remnant, we assume conservation of $E$, $J$ and $M_{\rm T}\equiv M_{\rm WD}+M_{\rm NS}$. 
We further assume that after tidal disruption all of the WD debris forms the envelope surrounding
the point mass NS. 
Accordingly, Eqs.~\eqref{Trot} and \eqref{Wtot} yield
\labeq{ToW2}{
\frac{T}{|W_{\rm t}|}=\frac{5 J^2 \lambda}{4\kappa_n k_5 M_{\rm WD}^3 Rg(\lambda)}, 
}
where the envelope is assumed to have a polytropic index $n$, constant $K$ and volume radius $R$. 
Angular momentum conservation amounts to substituting $J^2$ in Eq.~\eqref{ToW2} from Eq.~\eqref{Jinit2}. We then find
\labeq{ToW3}{\begin{split}
\frac{T}{|W_{\rm t}|} = & \frac{5\lambda}{4\kappa_n k_5g(\lambda)q(q+1)f(q)} \\
					& \hspace{1cm}\times \bigg[1+\frac{2\kappa_{n'}(1+q)f^2(q)}{5}\bigg]^2\bigg(\frac{R_{\rm WD}}{R}\bigg), 
\end{split}
}

We now impose energy conservation. This amounts to setting the RHS of Eq.~\eqref{Eeq} equal to the RHS of Eq.~\eqref{Einit}.  Combining the resulting equation with Eq.~\eqref{Wtot}  we obtain
\labeq{RwdoR}{
\frac{R_{\rm WD}}{R}=\frac{3(1-\Delta)}{(3-n')k_5g(\lambda)}\frac{\bigg(\displaystyle \frac{3-n'}{5-n'}+\frac{f(q)}{2q}\bigg)}{\displaystyle\bigg(1-\frac{3-2n}{3-n}\frac{T}{|W_{\rm t}|}\bigg)}.
}

Equations  \eqref{ToW}, \eqref{ToW3}, and \eqref{RwdoR} form a system of three algebraic equations for the three unknowns $T/|W_{\rm t}|$, $\lambda$ and $R/R_{\rm WD}$. After we specify $n'$ and $K'$ (see below for cases considered in this work) we 
 solve this system via an iterative Newton-Raphson scheme. 
Once we obtain $T/|W_{\rm t}|$, $\lambda$ and $R/R_{\rm WD}$ we can determine all properties of the hot remnant as follows: Given the mass of the envelope $M_{\rm WD}$, we find the total potential energy $W_{\rm t}$, the central density $\rho_{\rm c}$ and the volume radius of the envelope from Eqs.~\eqref{Wtot}, \eqref{Mrhoc} and \eqref{Req}, respectively. Also, given $T/|W_{\rm t}|$ and $W_{\rm t}$, we find $T$ and in turn we calculate the angular velocity of the rotating remnant as
\labeq{Omegafinal}{
\Omega=\sqrt{\frac{2T}{I}},
}
where $I$ is the remnant's moment of inertia, given by Eq.~\eqref{Imom}. 

Furthermore, given the initial $K'$, $n'$ and $M_{\rm WD}$ we calculate the final polytropic constant $K$ as follows: The radius of the initial WD is given by Eq. \eqref{Rpolytrope}, except for constants $K'$ and $n'$, 
\labeq{Rwdinitial}{
R_{\rm WD}=b_{n'}\bigg[\frac{(n'+1)K'}{4\pi}\bigg]^{n'/(3-n')}\bigg(\frac{M_{\rm WD}}{4\pi}\bigg)^{(1-n')/(3-n')}.
}
Thus, the ratio of $R_{\rm o}$ to $R_{\rm WD}$ is
\labeq{RwdoRo}{
\frac{R_{\rm o}}{R_{\rm WD}}=\frac{b_n}{b_{n'}}\bigg(\frac{M_{\rm WD}}{4\pi}\bigg)^{\beta(n)-\beta(n')}\bigg(\frac{1}{4\pi}\bigg)^{\gamma(n)-\gamma(n')}\delta(n,n'),
}
where 
\labeq{miscdefinitions1}{
\beta(n)= \frac{1-n}{3-n} \mbox{ \ , \ } \gamma(n)=\frac{n}{3-n},
}
and
\labeq{miscdefinitions2}{
\delta(n,n') = \frac{(n+1)^{\gamma(n)}}{(n'+1)^{\gamma(n')}}\frac{K^{\gamma(n)}}{K'^{\gamma(n')}}.
}
In the limiting case $n=n'$ Eq.~\eqref{RwdoRo} reduces to 
\labeq{}{
\frac{R_{\rm o}}{R_{\rm WD}}=\bigg(\frac{K}{K'}\bigg)^{\gamma(n)}.
}
Therefore, once we determine $R/R_{\rm WD}$, Eqs. \eqref{Req} and \eqref{RwdoRo} form one equation which we use to determine $K$ (which is related to the final entropy). 

\begin{center}
\begin{table*} [t]
\caption{WDNS binaries from Tables \ref{table1} and \ref{table2} which will undergo UMT. The table shows the properties of the initial WD and those of the hot remnant, for which we always assign $n=1.5$. The initial WD is described as an $n'=1.5$ (low-mass) or an $n'=2.9$ (high-mass) polytrope. The columns from left to right give the name of the object, the central density of the initial WD, the radius of the initial WD, the central density of the envelope of the remnant, the ratio of the final (volume) radius to the initial WD radius, the ratio of the spin kinetic energy to the total gravitational potential energy of the remnant, the dimensionless angular momentum of the remnant (in geometrized units), the spin angular velocity of the remnant and the central temperature of the envelope.}
\begin{tabular*}{1.0\textwidth}{@{\extracolsep{\fill}} l c c c c c c c c }\hline
\multicolumn{9}{c}{$n'=n=1.5$} \\ \hline\hline
PSR  &  $\rho_{\rm c}^{0}(10^6 {\rm g/cm}^{3})$ & $R_{\rm WD}({\rm Km})$  &  $\rho_{\rm c}^{e}$ $(10^5 {\rm g/cm}^{3})$ & $R/R_{\rm WD}$  &  $T/|W_{\rm t}|$   & $J/M_{\rm T}^2$   & 
 $\Omega (s^{-1})$  &  $\Theta_{\rm c}  (10^9 \rm K)$   \\
\hline 
B2303$+$46   	& $6.88$   &  $8136$       &   $5.11$      &    $2.38$             &    $ 0.26$      &   $19.7$      & $9.22$         & $1.46$                        \\  \hline
J1157$-$5114   	& $5.29$   &  $8500$       &   $3.26$     &     $2.53$             &    $ 0.27$      &   $20.8$      & $7.74$         & $1.26$                         \\ \hline
J1141$-$6545    	& $4.23$   &  $8821$       &   $2.44$      &    $2.59$             &    $ 0.27$      &   $22.0$      & $6.82$         & $1.10$                       \\  \hline
J1435$-$60    	& $4.92$   &  $8602$       &   $2.76$      &    $2.61$             &    $ 0.27$      &   $20.8$      & $7.30$         & $1.23$                       \\  \hline
J1453$-$58     	& $4.66$   &  $8682$       &   $2.51$      &    $2.65$             &    $ 0.27$      &   $21.0$      & $7.04$         & $1.19$                          \\  \hline
J1022$+$1001    & $3.09$   &  $9295$       &   $1.26$      &    $2.91$             &    $ 0.28$      &   $22.2$      & $5.36$         & $0.96$                         \\  \hline
B0655$+$64   	& $2.70$    &  $9510$      &   $0.99$     &     $3.01$             &    $ 0.29$      &   $22.5$      & $4.89$         & $0.90$                 \\  \hline
\multicolumn{9}{c}{$n'=2.9, \ n=1.5$} \\ \hline\hline 
 PSR  & $\rho_{\rm c}^{0}(10^8 {\rm g/cm})$ & $R_{\rm WD}({\rm Km})$  &  $\rho_{\rm c}^{e}$ $(10^5 {\rm g/cm})$ & $R/R_{\rm WD}$ & $T/|W_{\rm t}|$  & $J/M_{\rm T}^2$  & 
$\Omega (s^{-1})$  &  $\Theta_{\rm c}  (10^9 \rm K)$   \\ \hline
B2303$+$46   	& $3.38$   &  $4398$       &   $1.61$      &    $6.53$             &    $ 0.12$      &   $14.0$      & $4.52$         & $1.67$                        \\  \hline
J1157$-$5114   	& $0.68$   &  $7202$       &   $0.31$     &     $6.62$             &    $ 0.13$      &   $18.6$      & $2.14$         & $0.97$                         \\ \hline
J1435$-$60    	& $0.54$   &  $7685$       &   $0.24$      &    $6.67$             &    $ 0.14$      &   $19.1$      & $1.96$         & $0.91$                \\  \hline\hline 

\end{tabular*}
\label{table5}
\end{table*}
\end{center}

Finally, given $K$ we also obtain an estimate for the temperature of the envelope $\Theta$, if we approximate the total pressure $P$ as the sum
of the cold pressure $P'$ and thermal pressure $P_{\rm th}$. This gives the correct form in the extreme cold (degenerate) and extreme hot (Maxwell-Boltzmann) limits. 
Then the thermal pressure is
\labeq{Pth1}{
\frac{P_{\rm th}}{\rho} = \frac{P}{\rho}-\frac{P'}{\rho}=K \rho^{1/n}-K'\rho^{1/n'}.
}
The temperature is then 
given by
\labeq{Pth2}{
\Theta=\frac{{\mu m_u}}{k}\frac{P_{\rm th}}{\rho}.
}
Here $m_u$ is the atomic mass unit and $\mu$ is the mean molecular weight, which, for fully ionized $^{12}_{\ 6}C$ material,  has the value $\mu=12/7$ 
and $k$ Boltzmann's constant.
 As noted at the beginning of Section \ref{tidal disruption}, we will always assume an adiabatic index $\Gamma=5/3$ ($n=3/2$) for the hot remnant. 

In Table \ref{table5}  we list the properties of WDNS remnants in the hot scenario by modeling those binaries from Tables \ref{table1} and \ref{table2} which will eventually undergo UMT. We have considered two different models: (a) The initial WD is described by  $n'=1.5$, and (b) the initial WD is described by $n'=2.9$.

For case (a) we set $K'=3.161\times10^{12} (cgs)$, which corresponds to the nonrelativistic limit of the EOS for an ideal degenerate electron gas. The data in the table shows that the volume radius of the hot remnant is always larger than the radius of the progenitor WD, i.e.~the hot remnant is puffed up. The spin frequency of the remnant is of order few ${\rm Hz}$ and $J/M_{\rm T}^2$ is of order $20$ and hence significantly larger than unity.  
This result implies that the remnant cannot collapse to form a Kerr black hole unless some process either removes angular momentum from the system, or forms a massive disk about a hole. Also, the ratio $T/|W_{\rm t}|$ is almost always comparable to $0.27$. This implies that these configurations are marginally prone to the development of the bar mode instability on a dynamical timescale \cite{ChandrasekharEllips, LRS93a, ShibataBarmode, SaijoBarmode}. 

The initial and final densities we calculated are consistent with the choice of polytropic EOS we made, because they are not much larger than $10^6 {\rm g/cm^3}$, below which a $n=1.5$ index describes well the cold degenerate matter \cite{Shapiro}. However, note that the WD components of objects B2303$+$46, J1157$-$5114 and J1435$-$60 are large enough ($> 1.1 M_\odot$) that the degenerate electrons may be relativistic. If this is so, a polytropic index $n'\approx 3$ better describes the WD EOS. 

To check this we  used our $\rho_{\rm c}$ - $M_{\rm WD}$ data from the integration of the TOV equations in conjunction with the EOS, described in Section \ref{eq of state}. We find that the central densities which correspond to the masses of the WD component of objects B2303$+$46, J1157$-$5114 and J1435$-$60 are $3.38\times 10^8 {\rm g/cm}^{3}$, $6.75\times10^7 {\rm g/cm}^{3}$ and $5.36\times10^7 {\rm g/cm}^{3}$, respectively. These densities are significantly larger than $10^6  {\rm g/cm}^{3}$, and hence these high mass WDs are best studied by a polytropic index $n'\approx 3$. 

\begin{table*}[t]
\caption{The cold scenario for WDNS binaries of Table \ref{table5}. We show the properties of the cold remnant for $n'=n=1.5$ and  $n'=n=2.9$ (for the highest WD masses only). The columns from left to right give the name of the object, the central density of the envelope of the remnant, the ratio of the final (volume) radius to the initial WD radius, the ratio of the spin kinetic energy to the total gravitational potential energy of the remnant and the spin angular velocity of the remnant. The central density and radius of the initial WD and the dimensionless angular momentum of the remnant are the same as those listed in Table \ref{table5}.}
\begin{tabular*}{0.75\textwidth}{@{\extracolsep{\fill}} l c c c c  }\hline
\multicolumn{5}{c}{$n'=n=1.5$} \\ \hline\hline
PSR  &  $\rho_{\rm c}^{e}$ $(10^6 {\rm g/cm}^{3})$ & $R/R_{\rm WD}$  & $T/|W_{\rm t}|$  & $\Omega (s^{-1})$ \\
\hline 
B2303$+$46   	& $3.72$   &  $1.23$       &   $0.36$      &    $21.8$                           \\  \hline
J1157$-$5114   	& $2.60$   &  $1.27$       &   $0.38$     &     $18.5$                            \\ \hline
J1141$-$6545    	& $2.01$   &  $1.28$       &   $0.38$      &    $16.4$                          \\  \hline
J1435$-$60    	& $2.31$   &  $1.29$       &   $0.38$      &    $17.6$                          \\  \hline
J1453$-$58     	& $2.14$   &  $1.30$       &   $0.39$      &    $17.0$                             \\  \hline
J1022$+$1001    & $1.25$   &  $1.35$       &   $0.40$      &    $13.2$                            \\  \hline
B0655$+$64   	& $1.04$   &  $1.37$       &   $0.41$     &     $12.1$                    \\  \hline
\multicolumn{5}{c}{$n'=n=2.9$} \\ \hline\hline 

PSR  &  $\rho_{\rm c}^{e}$ $(10^7 {\rm g/cm}^{3})$ & $R/R_{\rm WD}$   &  $T/|W_{\rm t}|$  & $\Omega (s^{-1})$   \\\hline 
B2303$+$46   	& $5.83$   &  $1.80$       &   $0.35$      &    $66.9$                           \\  \hline
J1157$-$5114   	& $1.01$   &  $1.89$       &   $0.36$     &     $28.5$                            \\ \hline
J1435$-$60    	& $0.74$   &  $1.93$       &   $0.37$      &    $24.9$                          \\  \hline\hline 

\end{tabular*}
\label{table6}
\end{table*}

 We choose $n'=2.9$ to model these quasi-relativistic degenerate cases. This value is close to that of an extreme relativistic ideal degenerate electron gas and at the same time avoids singularities that arise for $n'=3$. Also, to have a polytropic EOS $P=K'\rho^{(1+1/n')}$ which is consistent with $\rho_{\rm c}$ and $M_{\rm WD}$, we choose $K'$ as follows: Using the central densities $3.38\times 10^8 {\rm g/cm}^{3}$, $6.75\times10^7 {\rm g/cm}^{3}$ and $5.36\times10^7 {\rm g/cm}^{3}$ of the WDs and the corresponding masses we compute a consistent $K'$ for each of these objects separately, using Eq. \eqref{polytropeM} and setting $n'=2.9$.

The last three rows of  Table \ref{table5} correspond to $n'=2.9$. The volume radius of the hot remnant is again
larger than the radius of the progenitor WD, but in this case the hot remnant is more puffed up than in the $n'=n=1.5$ case. The spin frequency of the remnant is of order few ${\rm Hz}$, but less than the corresponding $n'=n=1.5$ case. Furthermore, $J/M_{\rm T}^2$ remains much larger than unity, hence such remnants cannot collapse to form a Kerr black hole, unless some process removes angular momentum or forms a massive disk about a hole. In contrast to the $n'=n=1.5$ case, the  ratio $T/|W_{\rm t}|$ is less than $0.27$, but close to the secular instability limit for bar formation \cite{ChandrasekharEllips, LRS93a, ShibataBarmode, SaijoBarmode}.

\subsubsection{The ``cold'' remnant}

We determine the properties of cold remnants as follows: First, we assume that angular momentum is conserved and that the hot remnant cools by radiating the excess energy until $K'=K$ and $n'=n$, thereby returning to its original (degenerate) state. 
Equations  \eqref{ToW}, \eqref{Req} and \eqref{ToW3} form a system of three equations for the three unknowns $T/|W_{\rm t}|$, $\lambda$ and $R/R_{\rm WD}$. This is so, because in Eq. \eqref{Req} $R_{\rm o}=R_{\rm WD}$, which follows from Eq. \eqref{RwdoRo} under the assumption that $K'=K$ and $n'=n$. We solve this system of algebraic 
equations via an iterative Newton-Raphson scheme. Once we obtain $T/|W_{\rm t}|$, $\lambda$ and $R/R_{\rm WD}$, we determine all properties of the hot remnant as follows: Given the mass of the envelope $M_{\rm WD}$ and $n$, using Eqs. \eqref{Wtot}, \eqref{Mrhoc} we calculate the total potential energy $W_{\rm t}$ and the 
central density $\rho_{\rm c}$ of the envelope respectively. Finally, given $T/|W_{\rm t}|$ and $W_{\rm t}$, we find $T$ and in turn calculate the angular velocity of the rotating remnant via Eq. \eqref{Omegafinal}.

In Table \ref{table6} we list the properties of WDNS remnants in the cold scenario by considering the same binary systems as those listed in Table \ref{table5}. 
Just like in the hot scenario we consider two cases (a) $n'=n=1.5$ and  (b) $n'=n=2.9$.

 For $n'=1.5$ we again set $K'=K=3.161\cdot10^{12} (cgs)$. Note that the central densities are consistent with the choice of polytropic EOS we made. The data in the table shows that the volume radius of the cold remnant is 
always slightly larger than the radius of the progenitor WD, i.e. the remnant is slightly puffed up. The spin frequency of the 
remnant is of order of a few tens of ${\rm Hz}$ and $J/M_{\rm T}^2$ is of order $20$.  Therefore, these objects, too, cannot collapse to form a Kerr black hole unless some process removes angular momentum from the system or forms a massive disk about a hole. The ratio $T/|W_{\rm t}|$ is always significantly larger than $0.27$, indicating that these cold configurations are prone to the development of bars on a dynamical timescale.

Just like in the hot case, for the three highest mass WDs we also use $n'=2.9$ and $K'=K$ calculated as described in the previous section.  We list the results of this analysis in the last three rows of Table \ref{table6}. Note that the central densities of the cold envelope are marginally large enough to justify the choice $n=2.9$ perhaps with the exception of object J1435$-$60. The data in the table shows that the volume radii of the cold remnants are a little larger than the corresponding radii in the $n'=n=1.5$ case. The spin frequency is again of order of a few tens of ${\rm Hz}$ and larger than the corresponding values of Table \ref{table6}. These objects have precisely the same $J/M_{\rm T}^2$ as in the $n'=n=1.5$ case and hence, the remnants cannot collapse to form a Kerr black hole unless some process removes angular momentum from the system or forms a massive disk about a hole. Finally, the  ratio $T/|W_{\rm t}|$ is not very different than the $n'=n=1.5$ case. Thus, all these cold configurations are prone to the development of the dynamical bar mode instability. 

We emphasize again that all these estimates are tentative, and meant only as speculative previews of the possible characteristics of a WDNS remnant.
 Clearly, fully relativistic, dynamical simulations are needed to explore these scenarios conclusively.

\subsubsection{Alternative outcomes}

So far we have considered the possibility that the evolution of an unstable WDNS binary results in an equilibrium remnant with a NS at its center. However, this is not the only possible outcome. Alternatives include:
\begin{itemize}
\item Prompt collapse to a black hole immediately 
following NS merger with the disrupted WD;
\item Formation of a differentially rotating, hypermassive NS
following merger that ultimately undergoes delayed collapse to
a BH;
\item Ejection of appreciable WD debris, leaving a rotating NS 
remnant.
\end{itemize}

We point out that the situation here is 
reminiscent of the situation found in binary neutron star
mergers, where either prompt collapse to a BH or the
formation of a hypermassive star followed by delayed collapse
occurs \cite{2005PhRvD..71h4021S,2006PhRvD..73f4027S,2008PhRvD..78b4012L}.

The frequency of the gravitational radiation during the NS collapse phase is of order $\sqrt{G\bar\rho}$ (where $\bar\rho$ is the average NS density). If a bar is formed as the NS collapses the frequency of the GWs is of the same order as the spin frequency of the collapsing NS \cite{lrr-2003-2}. Hence, the GW signal from a collapsing NS could be between a few tens of ${\rm Hz}$ to a few ${\rm kHz}$, suitable for detection by Advanced LIGO. Therefore, just like in the UMT case with NS collapse, we could again have a scenario that begins with detectable LISA signals (from the WDNS inspiral), and ends with either a burst or quasiperiodic signal of GWs (from the NS collapse) detectable by ground-based interferometers. Note that the UMT case would have a  different GW signature than that of the SMT case. 
Therefore, these two scenarios are not degenerate and gravitational wave observations alone could discern which of the two scenarios took place.

Another interesting aspect of unstable WDNS binaries is that they might also be progenitors of gamma-ray bursts (GRBs). The material from the tidally disrupted WD may eventually form an accretion disk onto a BH remnant, if the NS collapses, whose lifetime could be of order typical GRB timescales.

On the other hand, if the lifetime of the disk is long enough and the conditions favorable for fragmentation to occur, WDNS mergers may offer a plausible 
route to the formation of planetary systems around isolated neutron stars. Currently, one such planetary system is known \cite{Wolszczan92} 
and the formation process of these systems remains an open question.

Fully relativistic hydrodynamics simulations are necessary to resolve the final fate of WDNS remnants. Although the field of numerical relativity has now matured enough to be able to handle problems involving BHBH and BHNS binaries, studying WDNS binary system offers, yet, an extra complication. This particular problem is extremely challenging  because it involves a NS of characteristic size $\sim10$ km and dynamical timescale $\sim 1$ ms and a WD of size $\sim10^3$ km and dynamical timescale $\sim 1$ s. Therefore, in order to be able to resolve all the different scales the full power of numerical relativity in conjunction with adaptive mesh refinement has to be employed. This is what we are now planning.

%
%

\section{Summary and Discussion} \label{summary}

In this paper we considered WDNS binaries and ``set the stage" for fully relativistic hydrodynamic simulations of the WDNS merger.
Like NSNS binaries, but unlike BHNS and BHBH binaries, WDNS binaries are known to exist and are abundant. We have compiled a list of 
observed WDNS binaries and their properties in Tables \ref{table1} and \ref{table2}.  

The emission of gravitational radiation will cause the orbital separation of a WDNS binary to shrink. 
We modeled corotational WDNS binaries in circular orbit and found that these models terminate at the Roche limit. At this point the binary can undergo either SMT and evolve on a secular timescale or UMT and evolve on a hydrodynamical timescale. 

Following the stability analysis of Verbunt and Rappaport \cite{Verbunt}, we showed that  the subsequent fate of the binary is determined by a critical mass ratio. Using the results of this stability analysis we predicted the possible fates of known WDNS binaries and we indicated this in our Tables~\ref{table1} and~\ref{table2}. Furthermore, based on population synthesis results by Nelemans et al. \cite{Nelemans01}, we gave an estimate of the number of LISA-resolved galactic binaries per year that will undergo SMT and the number of those that will undergo tidal disruption. We found that approximately up to $16$ WDNS binaries will undergo UMT (tidal disruption), and up to $20$ SMT.

In the case of SMT, the timescale of the mass transfer, and hence the timescale of the binary evolution, is set by the emission of gravitational radiation.
We treated this quasistationary SMT epoch by applying the approach of Clark and Eardley \cite{Clark77} and Faber et al. \cite{Faber}, also adopted in \cite{Rappaport82, Fryer99,Marsh}, and we estimated the corresponding gravitational waveforms. 

We also constructed approximate equilibrium configurations of rigidly rotating WDNS merged remnants in order to explore possible outcomes of WDNS mergers.  We found that unless some process removes angular momentum from the system or leads to the formation of a massive disk, these massive equilibrium configurations cannot collapse directly to form a Kerr black hole since their angular momentum ($J/M^2\sim 20>1$) exceeds the Kerr limit. Furthermore, in most of our case studies we found that the merged remnants have a ratio of spin kinetic to gravitational potential energy  greater than $0.27$, which implies that they are dynamically unstable to bar formation.  However, we emphasize that these preliminary results are at best approximate and must be confirmed and/or revised by detailed numerical simulations.

The fate of a merged WDNS binary depends on the initial mass of the cold progenitor stars, the 
degree of mass and angular momentum loss during the WD disruption and binary merger phases, the angular momentum profile of the 
WDNS remnant and the extent to which the remnant gas is heated by shocks as 
it pours onto the NS and forms an extended, massive mantle.
These issues all require a hydrodynamic simulation to resolve. 

Moreover, ascertaining whether or not the NS ultimately
undergoes a catastrophic collapse to a BH (either prompt or
delayed) requires that such a simulation be performed in full general relativity.  In fact, even the final fate of the NS in the alternative scenario in which there is a long epoch of SMT may also lead to catastrophic collapse, if the final NS mass exceeds the TOV mass limit. This scenario too will require a general relativistic hydrodynamic simulation to track. We plan to explore some of these alternative scenarios in detail in the future,
aided by simulations that employ our AMR relativistic hydrodynamics code.

\acknowledgments

It is a pleasure to thank R. Webbink, V. Kalogera and B. Willems for helpful discussions.  MM gratefully acknowledges support from the Maine Space Grant Consortium.  
This work was supported in part by NSF Grants PHY02-05155 and PHY06-50377 as well as NASA Grants NNG04GK54G and NNX07AG96G to the University of 
Illinois at Urbana-Champaign, and NSF Grant PHY07-56514 to Bowdoin College.

\appendix

\section{Accuracy of the approximate WD mass-radius relation \label{appA}}

In this appendix we demonstrate that Eq.~\eqref{Eggleton} (combined with Eq.~\eqref{EgglparamsOV}) is a good fit to the actual WD mass-radius data we obtained after numerically integrating the TOV equations adopting the EOS described in Section \ref{eq of state} for $\mu_e=2$.

\begin{figure*}[t]
\subfigure[{\label{fig:RMFITa}}]{\includegraphics[width=0.49\textwidth]{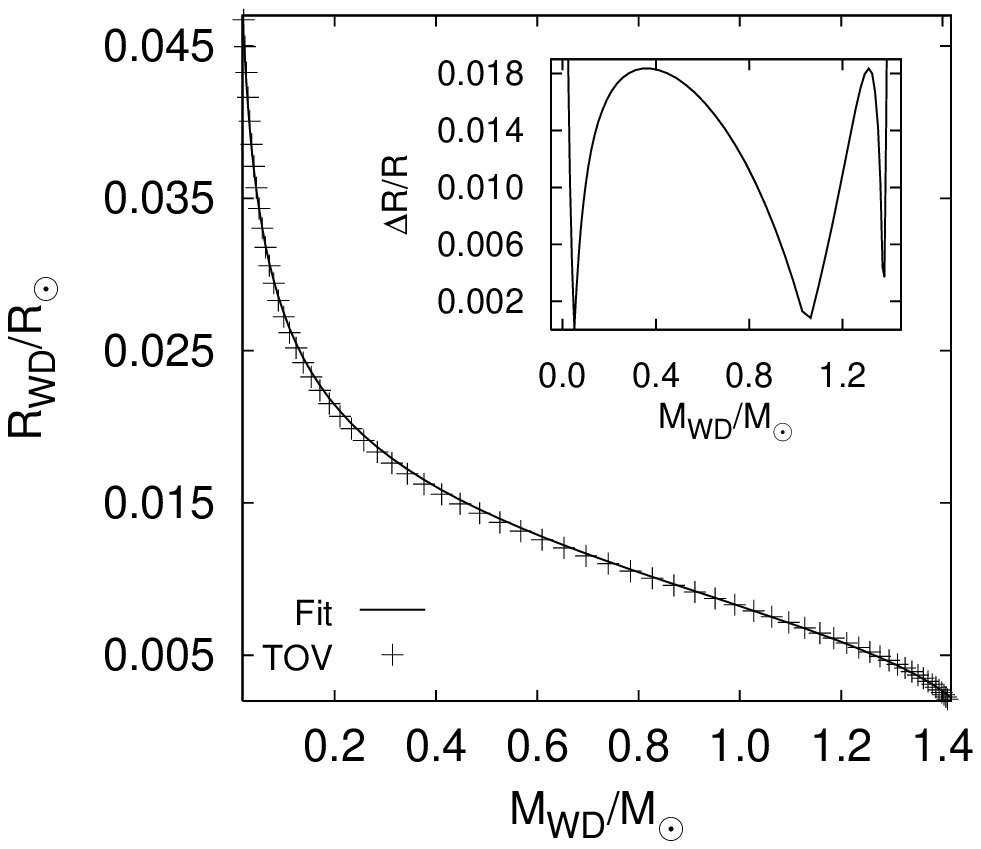}
}
\subfigure[{\label{fig:RMFITb}}]{\includegraphics[width=0.49\textwidth]{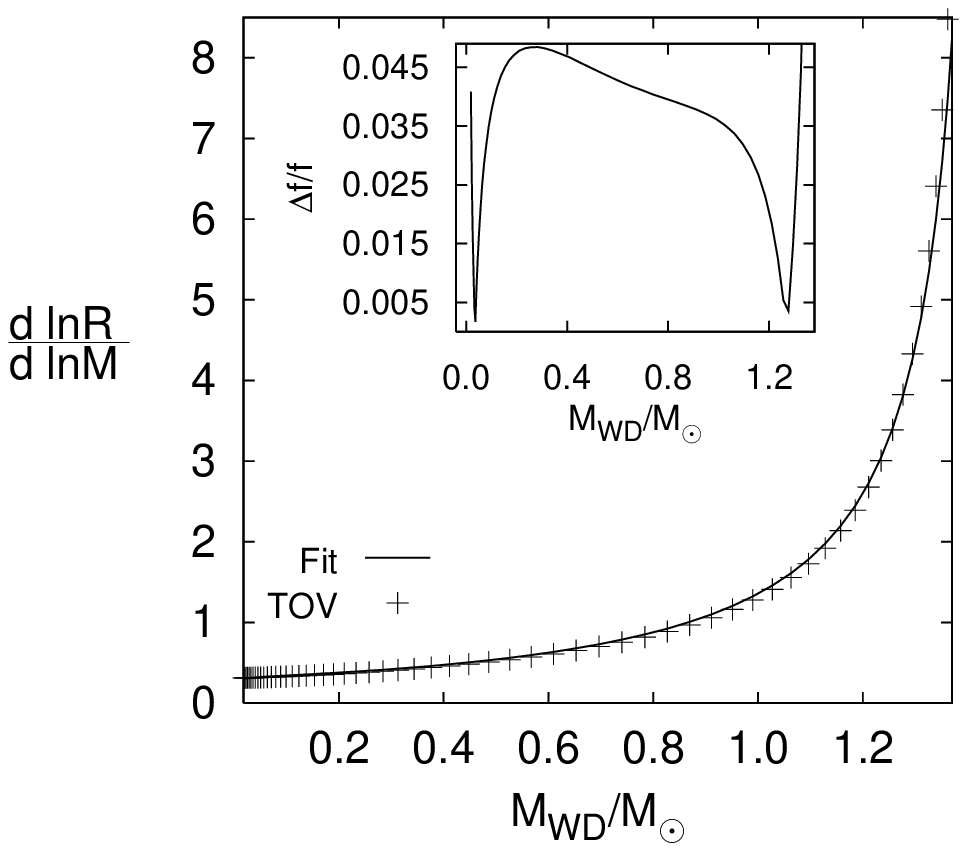}
}
\caption{{(a) Mass-Radius relation for a cold WD supported by an ideal degenerate electron EOS with $\mu_e=2$.  The inset shows the fractional error in the radius, $\Delta R/R$, where $\Delta R= |R_{\rm fit}-R_{\rm numerical}|$. (b) Logarithmic derivative of the WD radius. The inset shows the fractional error in the logarithmic derivative, $\Delta f/f$, where $\Delta f=|f_{\rm fit}-f_{\rm numerical}|$ and  $f=d\ln R/d\ln M$. ``TOV" means the data obtained by numerically integrating the TOV equilibrium equations and ``Fit" means the data obtained from Eq.~\eqref{Eggleton}, combined with Eq.~\eqref{EgglparamsOV}.
 \label{fig:RMFIT}
 }}
\end{figure*}

We find that the fractional error in the radius is less than $2\%$ for $M_{\rm WD}\in(0.02,1.39)M_\odot$ and the fractional error in the logarithmic derivative $d\ln R/d\ln M$ is less than $4.85\%$ for $M_{\rm WD}\in(0.02,1.33)M_\odot$ (see insets in Fig.~\ref{fig:RMFIT}, where we show the fractional errors). Note that the mass range over which the formula \eqref{Eggleton}  (combined with Eq.~\eqref{EgglparamsOV}) is accurate are those which are of astrophysical interest for WDNS binaries.  We also note our modified expression for the relativistic mass-radius relation represents an improvement over the original Eggleton approximation.

\bibliography{wdns}

\end{document}